# Reaching New Heights in Astronomy — ESO Long Term Perspectives

Tim de Zeeuw[1]

[1] ESO

A comprehensive description of ESO in the current global astronomical context, and its plans for the next decade and beyond, are presented. This survey covers all aspects of the Organisation, including the optical-infrared programme at the La Silla Paranal Observatory, the submillimetre facilities ALMA and APEX, the construction of the 39-metre European Extremely Large Telescope and the science operation of these facilities. An extension of the current optical/infrared/submillimetre facilities into multi-messenger astronomy has been made with the decision to host the southern Cherenkov Telescope Array at Paranal. The structure of the Organisation is presented and the further development of the staff is described within the scope of the long-range financial planning. The role of Chile is highlighted and expansion of the number of Member States beyond the current 15 is discussed. The strengths of the ESO model, together with challenges as well as possible new opportunities and initiatives, are examined and a strategy for the future of ESO is outlined.

## 1. Introduction

Astronomy is one of the oldest sciences and can be defined as the study of the Universe and everything in it. Advances in technology during the past half-century have made it possible to build increasingly powerful facilities on the ground and in space to study the Universe across the entire electromagnetic spectrum and to detect particles and gravitational waves from celestial sources. This has resulted in tremendous advances, including: the discovery of the mysterious dark energy; probing the extreme physics of black holes, supernovae and gamma-ray bursts; understanding the formation and evolution of stars and galaxies; and the direct detection and study of planets around other stars, which may have the potential to harbour life. These discoveries address some of the most fundamental questions in science, are of enormous interest to the general public, and are instrumental in stimulating young people to consider a career in science or engineering, which is important for our society.

The construction of a major astronomical facility typically takes a decade or longer. ESO has built three such observatories (on La Silla, Paranal and Chajnantor in Chile) since its founding in 1962 and is now constructing the European Extremely Large Telescope (ELT) on Armazones near Paranal. They take advantage of developments in technology and in turn stimulate further technology development. ESO's observatories make key contributions to all aspects of astrophysics and are the main ground-based observational resource for astronomers in most of the Member States.

### 1.1 The 2004 Council Resolution on Scientific Strategy

ESO's current programme is guided by the Council Resolution on Scientific Strategy, which was formulated in 2004 when ESO had 11 Member States. The Very Large Telescope (VLT) had entered full operations with all four Unit Telescopes (UTs), but not all of the first-generation instruments had been completed. An additional instrument, the High Acuity Wide field $K$-band Imager (HAWK-I), was in development, and four second-generation instruments had been selected (X-shooter, the $K$-band Multi Object Spectrograph [KMOS], the Spectro-Polarimetric High-contrast Exoplanet REsearch instrument [SPHERE] and the Multi Unit Spectroscopic Explorer [MUSE]). The VLT Interferometer (VLTI) was operational with the two-beam mid-infrared instrument MIDI. The VLT Survey Telescope (VST) was under construction but had experienced delays. The Visible and Infrared Survey Telescope for Astronomy (VISTA) had been agreed under the UK accession to ESO, and construction had started. ESO was operating three telescopes on La Silla and hosted a number of independent telescope projects there. The Atacama Large Millimeter/submillimeter Array (ALMA) was under construction by the partnership between East Asia, ESO and North America, but re-baselining was required to contain costs and place the main contracts. Phase A design studies for ESO's OWL concept and the competing EURO50 concept for extremely large optical-infrared telescopes were under development. Other giant telescope projects were gaining momentum elsewhere in the world. The ESO Council realised the need to develop a policy for the near future, which led to the Resolution on Scientific Strategy. The Resolution is included in full in the Appendix.

At the time of writing, twelve years later, ESO's programme at visual/infrared wavelengths has Paranal as its flagship, and the well-instrumented system comprising the VLT and VLTI, supplemented by VISTA and VST, is world-leading. Following extensive design studies, the 39-metre ELT is under construction on Armazones, as an integral part of the Paranal system, and is on track for first light in 2024. It has a good chance of being the first of the next generation of giant telescopes to go into operation. The venerable 3.6-metre telescope and the New Technology Telescope (NTT) on La Silla continue to produce excellent science. In the submillimetre regime, the Atacama Pathfinder Experiment (APEX) came on line in 2006 and the ALMA Partnership has completed the construction of the transformational array which is now operational on Chajnantor. In addition, ESO will host the Cherenkov Telescope Array (CTA) South in the Paranal area, and will operate it for the CTA Partnership on a cost-reimbursement basis.

ESO's internal organisation has been adapted to carry out this challenging portfolio of major programmes. Many internal processes have been streamlined and harmonised. Since early 2014 all staff at Headquarters work in a single set of interconnected buildings on the ESO premises and are no longer spread over multiple buildings on the Garching research campus. A world-leading outreach and education centre, the ESO Supernova Planetarium & Visitor Centre, donated by the Klaus Tschira Foundation, will be inaugurated in 2018.

In this same period, the accession of Spain (2006), the Czech Republic (2007), Austria (2009) and Poland (2015) increased the number of ESO Member

States to 15. The Brazilian accession agreement, signed in December 2010, was ratified by the Brazilian Congress and Senate in 2015, but still awaits the final signature by the President. Formal and informal discussions with a number of other countries regarding membership are ongoing.

The entire programme is under significant financial pressure caused by factors beyond ESO's control. These include rapidly increasing costs for labour in Chile, unfavourable exchange rates and special organisational contributions to the CERN Pension Fund.

1.2 Global context

State-of-the-art fully steerable optical/infrared telescopes with primary mirrors 8 metres, or larger, in diameter are operational at various locations around the world, supported by different organisations or partnerships. The main centre in the northern hemisphere is Mauna Kea, which hosts the two Keck telescopes, the Subaru telescope and the Gemini North telescope. The Large Binocular Telescope on Mount Graham and the Gran Telescopio Canarias on La Palma provide northern access to astronomers from (some of) the ESO Member States. The Gemini South telescope on Pachón in Chile and the VLT on Paranal cover the southern hemisphere. Two independent international partnerships aim to build the 25-metre Giant Magellan Telescope (GMT) on Las Campanas in Chile and the Thirty Meter Telescope (TMT). The latter was initially planned to be on Mauna Kea, but the location is presently in doubt. In the submillimetre radio regime ALMA has no competition (thanks to its sensitivity, baseline range, size and global character), except for the upgraded Plateau de Bure NOrthern Extended Millimeter Array (NOEMA), which can access the entire northern hemisphere.

Several new facilities will provide information complementary to that which ESO offers. The James Webb Space Telescope (JWST) will be launched in 2018, and will follow the Hubble, Spitzer and Herschel Space Telescopes as the new workhorse telescope in space. The European Space Agency (ESA) Gaia mission will provide very accurate positions and motions for more than a billion stars in the Milky Way by 2020. ESA's Euclid mission expects to provide an infrared map of the entire extragalactic sky by the middle of the next decade. The Large Synoptic Survey Telescope (LSST) on Cerro Pachón will start mapping the entire accessible sky twice per week in 2023, sensitive to new transient phenomena. At that time, the northern and southern components of the Cherenkov Telescope Array (CTA), designed to detect optical flashes in the atmosphere caused by high-energy gamma-rays from the Universe, should be operational as well. Ground-based facilities in the design phase, or on the drawing board, include the Square Kilometre Array, to be built in stages in South Africa and Australia, a large single submillimetre dish on Chajnantor, a large wide-field telescope for highly multiplexed spectroscopic surveys, new facilities to measure cosmic rays and a world-wide network of gravitational wave detectors with sufficient angular resolution to allow electromagnetic telescopes to locate and characterise the sources.

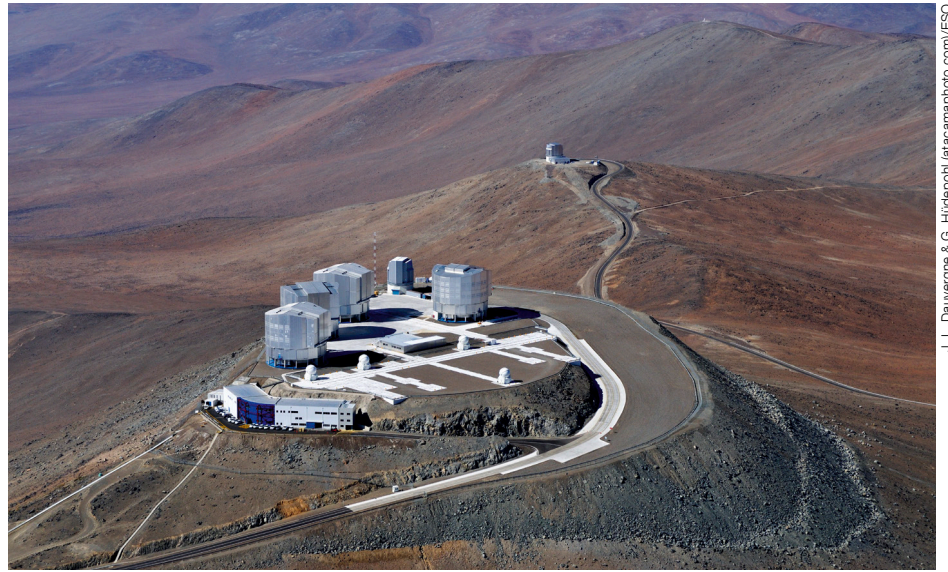

1.3 Long Term Perspectives

The approval of the ELT construction programme in December 2012, to which all Member States have committed significant additional funding despite the financial crisis of 2008, was followed two years later by the approval to split ELT construction into two phases. The first phase comprises 90% of the project. It has been authorised, is independent of the timing of the completion of the Brazilian accession, and was enabled by the accession of Poland. A significant uncertainty in the planning for the future was removed in June 2016 when Council gave approval to place the Phase 1 ELT contracts in accordance with first light in 2024, even if this might require making use of the credit facility in place with the European Investment Bank. It is therefore timely to present ESO's Long Term Perspectives, setting out the planned development of ESO's programme over the next fifteen years, and including the organisational perspective, financial planning and risks, options for funding of ELT Phase 2, opportunities beyond the baseline programme over a longer period, and an assessment of the ESO model.

Figure 1. Aerial view of the Paranal Observatory.

2. Optical-infrared programmes

ESO's programme started with the construction and operation of optical and infrared telescopes on La Silla, followed by the VLT system on Paranal, which is the current flagship, and will remain so for at least a decade beyond the period already foreseen in the Resolution on Scientific Strategy (Appendix). The next step in this wavelength regime is the construction of the ELT on nearby Armazones.

The Messenger 166 – December 2016    3

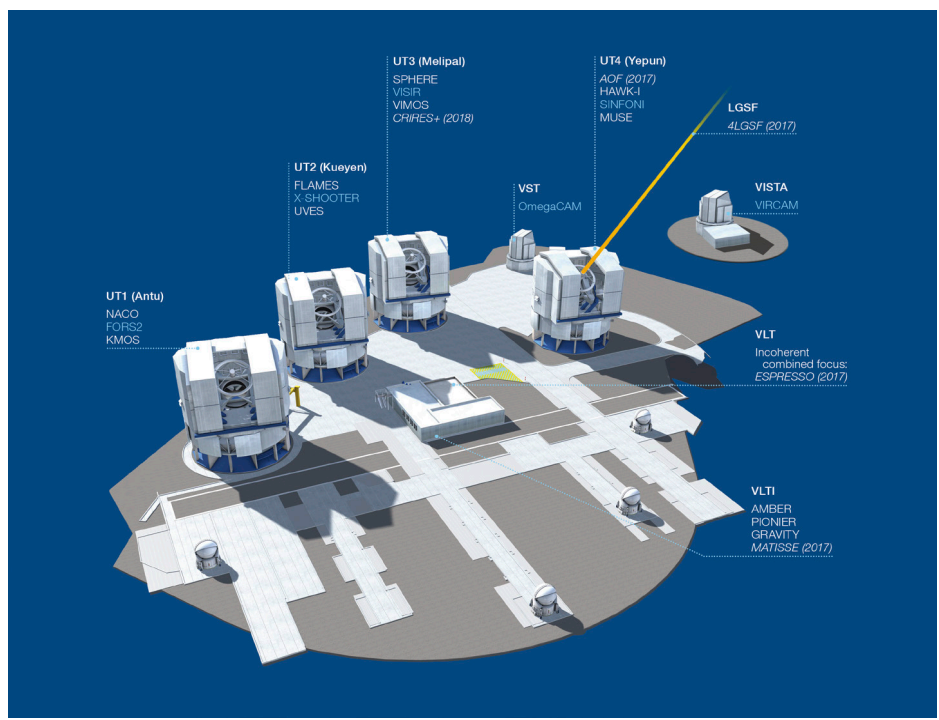

Figure 2. The Paranal system of telescopes and current instruments is shown, with the replacement instruments planned for 2017–2018 indicated (in italics).

It will be an integral part of the Paranal Observatory and will start observations by the middle of the next decade.

2.1 Paranal — the VLT system

The VLT was designed from the outset as an integrated system of four 8.2-metre UTs, including the possibility of combining the light collected by individual telescopes for optical interferometry (VLTI), either with the four UTs or with the four 1.8-metre Auxiliary Telescopes (ATs), providing superb high-angular-resolution capabilities. Most of the VLT and VLTI instruments are built in collaboration with consortia of scientific and technical institutes in the Member States, where ESO normally provides the hardware from its budget and the consortia provide the staff effort, for which they are compensated by an allocation of guaranteed time observations (GTO) with the instrument.

The VLT is currently equipped with 12 facility instruments mounted on the UTs and three VLTI facility instruments located in the coherent combined focus laboratory. VISTA and VST, each with a single dedicated camera, add wide-field imaging survey capabilities. A recent image of the VLT on Paranal is shown in Figure 1, with an annotated version of the telescopes and instruments in Figure 2. The complete system is unique among astronomical facilities world-wide.

The comprehensive coverage of parameter space in high-angular-resolution imaging and spectroscopy has led to: the detection and characterisation of planets and planetary systems orbiting other stars; the characterisation of circumstellar environments, protoplanetary discs, and stellar surfaces; high-resolution studies of the black hole in the Galactic Centre and of nearby galactic nuclei; measurements of the properties of the oldest stars; studies of interstellar and intergalactic matter; and exploration of the high-redshift Universe and early phases of galaxy formation.

The long-term instrumentation budget includes funding for a strategic programme of upgrades and new instruments every few years, maintaining the scientific capability of the Paranal system at the forefront by capitalising on developments in technology. This programme is summarised below, and is described in more detail in the Paranal Instrumentation Programme Plan[1].

2.1.1 Very Large Telescope

The mean lifetime of VLT instruments, set by technical and/or scientific obsolescence, was expected to be about ten years. In practice they have performed well for much longer, as upgrades have expanded their capabilities. The first generation of instruments was completed with the CRyogenic InfraRed Echelle Spectrometer (CRIRES) and HAWK-I in 2007. The four second-generation instruments, selected in 2002, are all operational, with X-shooter commissioned in 2009, KMOS in 2013, and SPHERE and MUSE in 2014. Figure 3 shows, for those VLT and VLTI instruments with spectroscopic capabilities, the plane of spectral resolving power and wavelength. Figure 4 displays the range of angular resolution vs. wavelength for the imaging modes of VLT and VLTI instruments.

In 2013 the PARSEC laser guide star was replaced by PARLA, which uses a fibre laser developed by ESO in collaboration with industry and provides a large increase in operational efficiency. The Adaptive Optics Facility (AOF) equips UT4 with four powerful sodium lasers (the 4 Laser Guide Star Facility, or 4LGSF) and a 1170-actuator deformable secondary mirror to enable diffraction-limited imaging and spectroscopy. In addition, the AOF contains two wavefront-sensor systems (the Ground Atmospheric Layer Adaptive optiCs for Spectroscopic Imaging [GALACSI] and the GRound layer Adaptive optics Assisted by Lasers [GRAAL]) to provide users with optimised adaptive optics (AO) modes with the MUSE and HAWK-I instruments, respectively. GRAAL was installed and tested at the telescope in 2015. The 4LGSF was commissioned on UT4 in April 2016 (see Figure 5), and the deformable secondary saw first light in October 2016. GALACSI will follow in early 2017.

The VLT Imager and Spectrometer for mid-InfraRed (VISIR) was upgraded in 2015/16 with new detectors and instrument modes and is now back in science operation with imaging and spectroscopy, coronagraphy, burst mode and sparse aperture mask modes all commissioned.

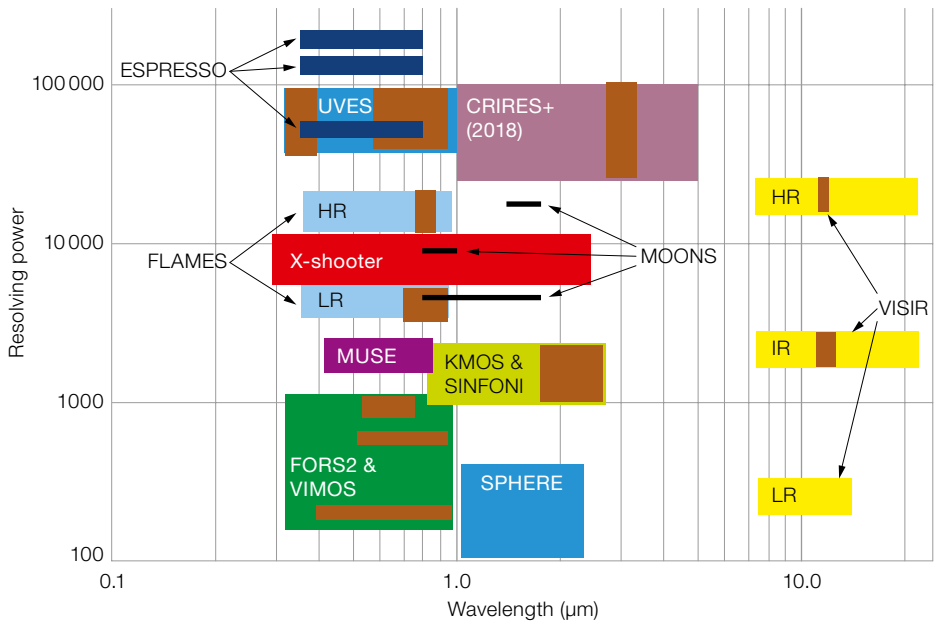

Figure 3. The spectral resolving power versus wavelength charted for current and planned VLT and VLTI spectroscopic instruments (brown areas designate higher resolution modes).

The next instrument to reach Paranal will be the Echelle SPectrograph for Rocky Exoplanet and Stable Spectroscopic Observations (ESPRESSO). It will take high-stability high-resolution spectroscopy to a new level and will be able to use any of the four UTs, or combine the light collected by all UTs simultaneously to observe an object with a 16-metre-equivalent telescope. This requires the commissioning of the four coudé trains, which is nearly completed, followed by commissioning of the instrument itself in 2017.

The upgrade of CRIRES to a cross-dispersed high-resolution near-infrared spectrograph (CRIRES+) covering the entire 1–5 µm wavelength range is on track for completion in 2018.

The Multi Object Optical and Near-infrared Spectrograph (MOONS), with wide-field and 0.8–1.8 µm coverage, is in development. It will be fibre-fed, and will have at least 800 (with a goal of 1000) fibres over a total field of 25 arcminutes in diameter. There will be two spectral resolutions: ~ 4000 spanning the full wavelength range and a higher-resolution mode which gives ~ 9000 in the I-band and ~ 20 000 in a region of the H-band. First light is planned for 2020.

In 2013 a Phase A study was completed for the Cassegrain U-band Brazil-ESO spectrograph (CUBES), with a strong role for Brazilian institutions. CUBES will go into development once the Brazilian ratification procedure is complete. If this should fail, other avenues could be explored.

NAOS-CONICA (NACO) was moved to UT1 in 2013 to make room for MUSE on UT4. NACO is aging and steps have been taken to allow it to continue in operation until it can be replaced by ERIS (the Enhanced Resolution Imaging Spectrometer). ERIS is a new instrument for the Cassegrain focus of UT4, consisting of a diffraction-limited infrared imager, an AO wavefront-sensor module, which uses the AOF deformable secondary mirror and any one of the four AOF lasers (for single-conjugate adaptive optics), and an upgraded version of the SPectrometer for Infrared Faint Field Imaging (SPIFFI), part of the Spectrograph for INtegral Field Observations in the Near Infrared (SINFONI), adapted to the new AO module. First light is planned for 2020.

Input for the selection of new instruments is provided via the Scientific Technical Committee (STC) and its sub-committees, at scientific conferences, or directly by the community. The emphasis is on 8-metre telescope science, rather than on technological concepts. Two recent workshops of scientists and AO experts focused on the definition of a new AO instrument in order to achieve the best possible scientific impact in an era of continuing adaptive optics developments on other telescopes and the expected availability of the James Webb Space Telescope (see Leibundgut et al. p. 62 and the Report to the STC[2]). One such idea foresees a relatively wide-field multi-conjugate AO (MCAO) imager, pushing bluewards of the traditional near-infrared MCAO regime, towards the peak of most stellar blackbodies in the optical.

The Breakthrough Foundation is interested in expediting the search for exoplanets through support of innovative approaches. It will support an experiment to search in the mid-infrared for planets orbiting in the Alpha Centauri system, by using VISIR in conjunction with the AOF. This is envisaged for 2019, and will also allow the testing of promising technologies that will benefit instruments on the ELT.

Other ideas are being developed in the community, including, for example, using SPHERE coupled with ESPRESSO for exoplanet characterisation. Other possibilities for the future are described in the STC report Paranal in the Era of the ELT[3].

2.1.2 VLT Interferometer
The VLTI offers a unique and world-leading capability for high-angular-resolution observations in the near- and mid-infrared, using either the UTs or the ATs (see Figures 3 and 4). The contribution of the VLTI to specific areas of stellar and extragalactic astronomy dominates the science output of all optical/infrared interferometers worldwide.

The Phase Referenced Imaging and Micro-arcsecond Astrometry facility (PRIMA), intended to enable astrometry at 10-micro-arcsecond accuracy with the VLTI, was cancelled in 2015 owing to continued delays and the prospect that completion was still at least three years away. This meant that the opportunity had been lost to carry out the foreseen science application in a timely way, given the successful launch of the Gaia mission. After an external review, the STC endorsed the proposal to cease activities on PRIMA and concentrate future efforts on ensuring the success of the second generation VLTI instruments. The lessons of PRIMA were formally reviewed and are being fed back to all ESO projects.

The VLTI was taken off line for seven months in 2015 in order to put in place all the infrastructure upgrades and new infrastructure needed to support the second-generation four-beam instruments GRAVITY and the Multi AperTure mid-Infrared SpectroScopic Experiment (MATISSE). This has resulted in a new AT maintenance station on the VLT platform, a comprehensive upgrade of the VLTI laboratory, installation of star separators on all ATs and UTs and global performance improvements targeted at achieving the full performance of GRAVITY (Woillez et al., 2015). The ATs will be equipped with the New Adaptive Optics Module for Interferometry (NAOMI) in the near future, and use of the GRAVITY fringe tracker for MATISSE will be a precursor to the possible development of a second-generation fringe tracker.

The Precision Integrated Optics Near-infrared Imaging ExpeRiment (PIONIER) was originally a visitor instrument and is now available to the community. GRAVITY is close to the end of commissioning, and has already made the first observations of the Galactic Centre with an astrometric precision which equals that achieved by classical AO techniques. MATISSE is scheduled to arrive on Paranal in 2017.

Looking further ahead, the VLTI will continue to provide the highest angular resolution, even in the ELT era. The rising demand for very high-resolution imaging of stellar surfaces, close circumstellar environments and extragalactic sources suggests a path forward which includes completing GRAVITY, MATISSE and the second-generation fringe tracker, continuing to offer a visitor focus on the VLTI (once the Astronomical Multi-BEam combineR [AMBER] is decommissioned), and exploring six-telescope imaging capabilities with the existing infrastructure (if the required funding can be identified). This could be done either with a combination of ATs and UTs, or by adding two fixed ATs.

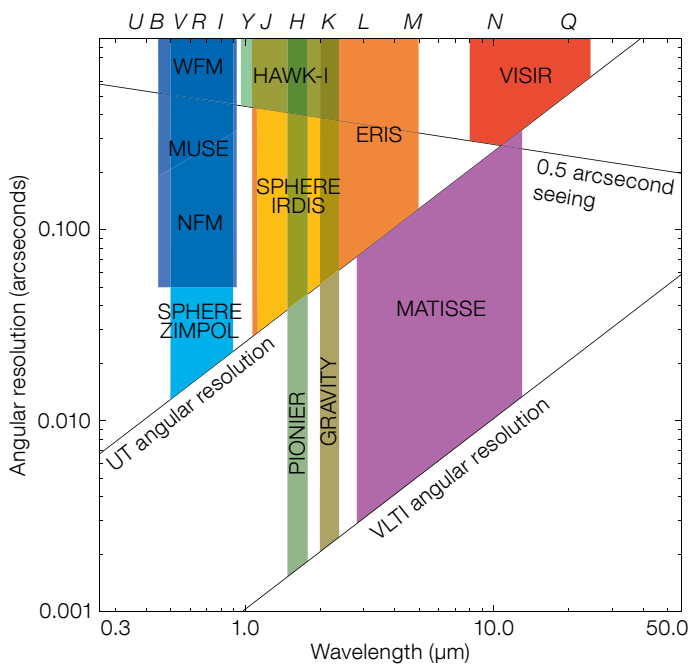

Figure 4. The angular resolution versus wavelength plane of the current and planned VLT and VLTI imaging instruments.

2.1.3 Survey telescopes

VISTA, equipped with the infrared camera VIRCAM, started scientific operation in early 2010. Six public survey imaging programmes, together requiring at least five years of observing time, are currently nearing completion and are already highly cited. Seven new imaging surveys, selected following peer review, are due to commence in 2017.

VST, equipped with OmegaCAM, started scientific operation in October 2011 and public and guaranteed-time surveys are being carried out. VST will be operational at least through 2021 with an increasing fraction of GTO time.

The public imaging surveys include studies of the entire sky accessible to Paranal, as well as more focused studies of the Milky Way, the Magellanic Clouds and deep extragalactic fields. These surveys are conducted by international teams, together with data centres in the Member States, coordinated by ESO.

VISTA is the ESO telescope with the largest field-of-view. In 2020/21 VIRCAM will be replaced by the 4-metre Multi-Object Spectroscopic Telescope (4MOST) instrument. With a field-of-view of more than three square degrees, 4MOST will host up to 2400 fibres, working in the optical (0.3–0.9 μm) regime. The goal is to have 1600 fibres that feed two lower-resolution ($R \sim 5000$) spectrographs, and 800 fibres feeding one higher-resolution ($R \sim 18\,000$) spectrograph.

The competitiveness of the VST will need to be monitored over the coming years. There are several optical surveys planned or ongoing on more powerful telescopes elsewhere, which will cover larger sky areas and will go to greater depth than is possible with the VST. The strengths of the VST that could be exploited in the future are the image quality achievable at Paranal and the blue sensitivity of OmegaCAM. A further potentially interesting area would be narrow-band imaging. The VST agreement runs for ten years, and concludes in 2021. At present, there are no funds in the ESO budget for VST activities beyond 2021. This does not, however, preclude its becoming a non-ESO-operated hosted telescope.

Building on the public imaging surveys with VISTA and VST, teams of astronomers have recently been undertaking four extremely ambitious public spectroscopic surveys. The first of these exploits synergies with the ESA Gaia mission to provide the first homogeneous overview of the kinematics and elemental abundances throughout the Milky Way. Another, the Public ESO Spectroscopic Survey of Transient Objects (PESSTO), is building a more comprehensive understanding of the exotic, explosive Universe. The most recent surveys, VANDELS and the Large Early GAlaxy Census (LEGA-C), are using the wide-field VIsible Multi-Object Spectrograph (VIMOS) on UT3 to probe the physics in many thousands of galaxies, connecting galaxies in the early Universe with those in the present day. It is likely that even more ambitious spectroscopic surveys of stars and galaxies will be undertaken from the early 2020s in support of ESA's PLAnetary Transits and Oscillations of stars (PLATO) and Euclid missions.

2.1.4 Hosted telescopes

Since 2015 Paranal has hosted the Next Generation Transit Survey (NGTS), a small university-led robotic experiment with an array of twelve 20-cm telescopes coupled to large-format, deep-depletion CCD cameras with the prime objective of detecting transiting Neptune-size planets orbiting K and M stars. A second hosted telescope project, SPECULOOS (Search for habitable Planets EClipsing ULtra-cOOl Stars), is complementary to NGTS and will carry out a photometric survey designed to discover Earth-size planets transiting the brightest southern ultra-cool stars. SPECULOOS consists of four 1-metre robotic telescopes equipped with CCD cameras. The hosting of NGTS and SPECULOOS at Paranal is cost-neutral to ESO since all construction and running costs are reimbursed to ESO by the consortia. In return for hosting these projects, the ESO community receives high-level data products produced by NGTS and SPECULOOS that are available through the ESO Science Archive Facility (SAF).

Hosting NGTS and SPECULOOS on Paranal is an exception justified by the low water vapour content of the atmosphere, which allows these experiments to achieve the very high photometric precision required for characterising exoplanet transits. Otherwise the location of choice for robotic experiments and small telescope projects remains La Silla with its well-developed support infrastructure.

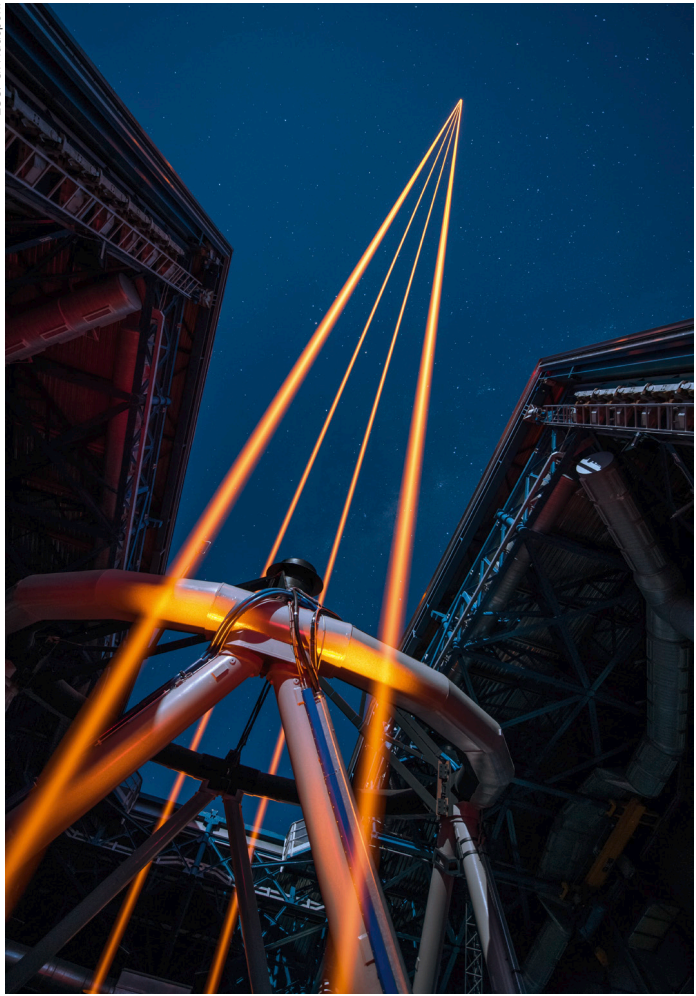

Figure 5. The four laser guide stars of the Adaptive Optics Facility on VLT Unit Telescope 4.

2.1.5 Obsolescence

The VLT started operations in 1999 and many of its sub-systems are ageing. The obsolescence of devices and components in these sub-systems puts at risk the availability of the telescope systems, scientific instruments and other operational equipment. VLT technology upgrades are carried out only after careful evaluation, and aim to use industrial standards and technologies with guaranteed long-term upgrade paths and maximum independence from control software choices. An additional consideration is to maximise common technologies between VLT and ELT in view of the integration of the ELT into the existing Paranal system.

The unavailability of electronics spares for the primary mirror (M1) cells of the UTs was identified early on as one of the highest risks for VLT operations. An obsolescence project was initiated in 2012 to replace the custom-made M1 cell electronics with commercial off-the-shelf electronics that ensures the long-term availability of spares. The first upgraded systems were installed and tested on UT1 in 2014 and deployment at the other UTs followed in 2015 and 2016.

In parallel, an obsolescence project to align the VLT emergency-stop and safety-chain system with evolved safety standards required upgrading existing installations with Siemens Safety programmable logic controllers (PLCs). UT2 was upgraded in 2013, followed by the other UTs from 2014 to 2016. The new PLC-based safety-chain system allowed seamless integration of the critical safety systems of the 4LGSF on UT4. An upgrade of the safety-chain system of VISTA is planned for 2017.

Regular operations of the 8-metre coating unit were stopped in 2013 because of the steadily decreasing quality of the aluminium coatings produced. After a year of intensive testing and analysis of the root cause of the problems, a contract was placed for the complete refurbishment of the coating unit. This was completed in 2016 with the recommissioning of the coating unit and the establishment of a rigorous long-term maintenance programme. The first production coating on a UT mirror in August 2016 was entirely successful.

2.1.6 Paranal infrastructure
A significant infrastructure upgrade that is being undertaken with the support of the Chilean government is the connection of the Paranal area to the Chilean electrical grid, via a 66 kV line. This runs from Paposo, 50 km to the south, to a new substation (see Figure 7), from which 23 kV distribution lines will connect to Paranal and Armazones. The connection will provide substantial savings in operational costs and allows ESO to take advantage of future developments in the provision of green energy to northern Chile. Completion is expected in the first half of 2017, perfectly timed for the major ramp-up of ELT construction activities on Armazones.

The arrival of the 23 kV line at Paranal requires an adaptation to the existing Paranal power distribution system at 10 kV and an upgrade of the Paranal power conditioning system to the expected characteristics of the grid power. Accordingly, a project was initiated in 2015 to install a combined flywheel/diesel generator system to provide simultaneously both power conditioning and backup power by early 2017, in time for the expected arrival of the grid connection. Operation and maintenance of the current multi-fuel turbine generator and procurement of liquified petroleum gas will be discontinued shortly thereafter.

No further major infrastructure upgrades are expected in Paranal in support of the VLT. All necessary additional infrastructure upgrades are driven by and financed through the Paranal Instrumentation Programme and the ELT Programme.

2.1.7 Science operations
ESO offers facilities for projects ranging from a few hours of observing time to surveys encompassing several hundred nights. The Paranal operational model has enabled new observing modes and opened monitoring and time-domain programmes that are hard to carry out in classical visitor mode. Service observing enables users to request rare observing conditions, with the rapid response mode allowing very swift access to an 8-metre telescope, while visitor mode allows the astronomer to make real-time decisions based on the data they have just acquired. Service mode is used approximately 70 % of the time, yet the fraction of time requested in this mode continues to grow, reaching 85 % in 2016. Thanks to the evolution of network bandwidth, reliability and security, additional possibilities can be envisioned, for instance remote participation in complex observations requiring real-time decisions. Such remote observing options are being studied, and could be deployed for some programmes and/or some telescopes. They would be introduced carefully, taking into account operational and financial constraints. An analysis of performance metrics collected since the start of VLT operations allows the linking of science operations and programme implementation to science return (Sterzik et al., 2015), which provides input for further fine-tuning of future integrated VLT and ELT operations.

The following key activities are being pursued with high priority:
– A comprehensive overhaul of the Phase 1 proposal submission and handling tools to remedy shortcomings revealed by the increased capabilities of the observing facilities and the complexity of proposed observing programmes;
– Provision of web-based Phase 2 observation preparation and execution tools for service and visitor mode programmes;
– Enhancements of the services offered by the SAF enabling users to comprehensively exploit ESO's data holdings so as to increase science return;
– Further optimisation of the allocation and scheduling of observations to ensure efficient operation of the AOF and of ESPRESSO using any, or all, incoherent UT foci;
– Integration of improvements in the reliability of weather forecasting into the short-term observing scheduling system.

An ESO-led community working group on time allocation will report in 2017. This working group has explored potential new approaches to proposal submission and peer review, including a critical look at the existing submission channels and response times, with a view to improving the synergies with other astronomical facilities, present and future, looking towards the ELT era.

All astronomical data obtained are routinely processed and rapidly made available to the user community through the SAF after quality control. Certified pipelines process the data stream for many of the instruments in unattended mode, removing the instrument signature, scaling to physical units and providing well-defined errors. A complementary stream of scientific data products is also produced by the community which greatly enhances the science archive via the Phase 3 submission process.

Managing multi-messenger, multi-wavelength, multi-facility data is an increasing challenge, yet is critical for tackling ever more complex science questions. ESO is ideally placed to meet these challenges, where open access to data is now widely recognised as an essential research infrastructure. A recent ESO-led community working group report on science data management (STC Report 580[4]) outlines how ESO should foster, coordinate and lead such collaborations, with activities directed internally and towards the community. The report recommends that ESO should maintain an active presence in activities relating to scientific data at a global level, and must also harness the expertise in its Member States, perhaps entering into collaborative arrangements with external parties for the delivery of specific data management functions.

2.2 Armazones — the ELT

The science enabled by the ELT focuses on three prominent areas: 1) detection and characterisation of exoplanets, with Earth-like planets directly accessible for the first time; 2) probing the formation and evolution of galaxies by enabling the study of resolved stellar populations in the full range of galaxies out to Virgo cluster distances, and in integrated light to high redshift; and 3) fundamental contributions to cosmology by measuring the properties of the first stars and galaxies, the nature of dark matter and dark energy, and a direct measurement of the acceleration of the Universe. These science cases require a 40-metre-class telescope, yielding higher sensitivity and angular resolution than GMT and TMT, and they benefit from a specific instrument suite and operational model.

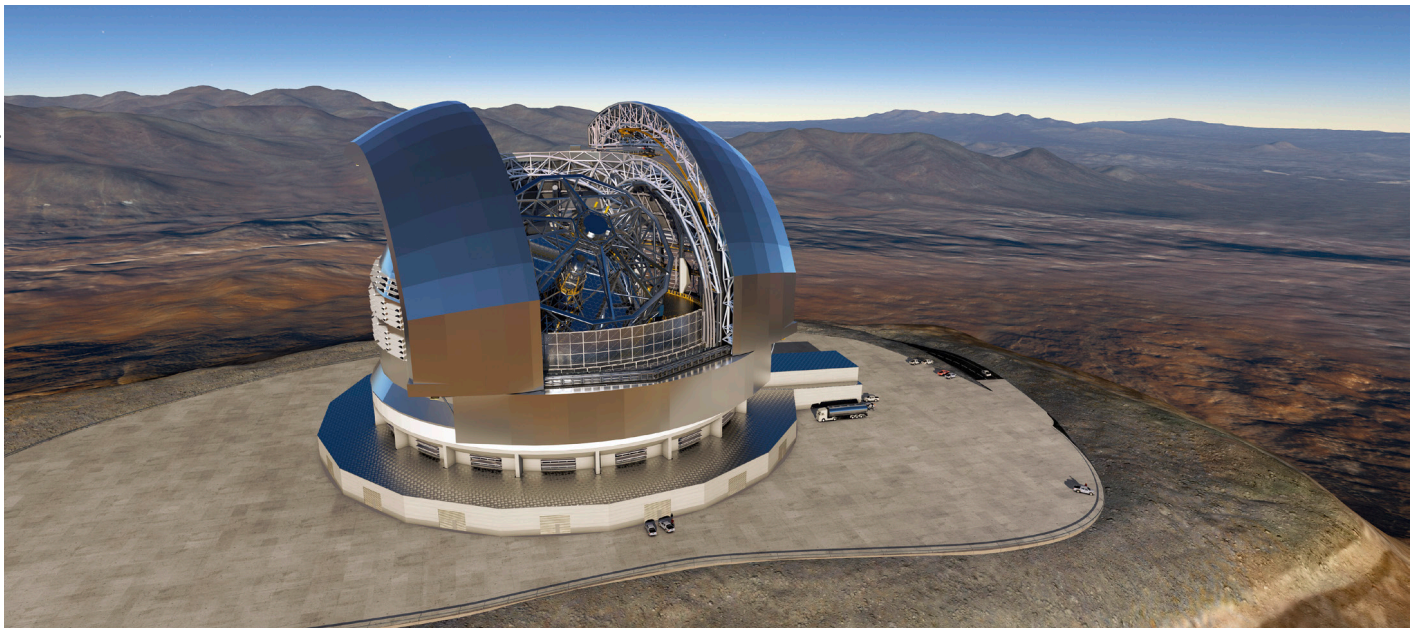

Figure 6. Artist's rendering of the European Extremely Large Telescope on Cerro Armazones in 2024.

2.2.1 Construction in two phases
In December 2014 Council authorised splitting the construction of the ELT into two phases. Phase 1 is for the 39-metre ELT with three instruments and an adaptive optics module, but without the laser tomography adaptive optics (LTAO) system and the five inner rings of segments of the primary mirror M1. It is affordable without Brazil as a Member State and Council has authorised the construction of this phase. Phase 2 will complete the baseline ELT by providing the missing LTAO module, the five inner rings of segments, additional segments for M1 coating and maintenance, and a second pre-focal station needed for additional instruments. The ELT operations budget will provide not only for on-site operations, but also for the ongoing instrumentation programme and major refurbishments of the telescope.

In June 2016 Council authorised placing the remaining contracts for the Phase 1 ELT on a schedule that leads to first light in 2024, even if this might require borrowing from the European Investment Bank during the peak of construction activities. This decision lowers cost and risk, and significantly increases the likelihood that the ELT will achieve first light before GMT and TMT, and be on sky simultaneously with JWST.

Council set the following constraints on the cost of ELT construction (in 2016 prices):
– The Phase 1 cost-to-completion is 1033 MEUR for first light in 2024, including 93 MEUR allocated for contingency and 124 MEUR for FTE (full time equivalent) costs;
– The Phase 2 cost-to-completion is 110 MEUR.

2.2.2 Phase 1
All activities are on track for first light in 2024. A new road to Armazones and the mountaintop platform for the telescope have been completed. Contracts for the final design and qualification of the M1 segment support (two parallel contracts for selection of the best design), for the design and fabrication of the mirror shell and support unit of the adaptive quaternary M4 mirror, as well as for the polishing of the 4.2-metre secondary mirror M2, are running. The very large contract for the Dome and Telescope Structure was signed in May 2016 and final design work is under way (Figure 6). The first stone is expected to be laid in mid 2017.

Since December 2014, much work has been done towards the procurement of the opto-mechanics (for example, blanks, polishing and supports for mirrors M1 to M5). The M1 polishing contract was approved by Finance Committee in November 2016 (the second-largest ELT contract after the Dome and Telescope Structure) together with the contracts for the M2 and M3 cells, and the edge sensors as well as the M2 and M3 blanks and M3 polishing. This brings the total material budget committed through external contracts and agreements to about 80 % of the ELT Phase 1 cost-to-completion, while leaving appropriate contingency. Other calls for tender are being prepared to allow contract signature in the next two years for the remaining equipment, including the blanks and polishing for M5, the blanks and the position actuators for the M1 segments, the serial manufacturing of the M1 segment supports, the laser guide star units, the pre-focal station, and the mirror washing and coating units.

The Phase 1 instruments are: the Multi-AO Imaging CAmera for Deep Observations (MICADO), a high-spatial-resolution multi-conjugate adaptive-optics-assisted camera/spectrograph; the High Angular Resolution Monolithic Optical and Near-infrared Integral field spectrograph (HARMONI), an adaptive-optics-assisted integral field spectrograph; and the Mid-infrared ELT Imager and Spectrograph (METIS). These instruments, and the Multi-conjugate Adaptive Optics RelaY (MAORY) coupled to MICADO, will be mounted on the Nasmyth A platform of the ELT. In accordance with the successful VLT model (section 2.1), agreements with consortia were put in place in 2015 for the design, fabrication and on-site installation of the instruments and the MAORY module. These commit ESO funding for the instrument hardware and secure significant additional expenditure for staff effort in the Member State institutions involved, compensated by the award of GTO to the consortia.

Phase A studies for the Multi-Object Spectrograph for Astrophysics, Intergalactic-medium studies and Cosmology (MOSAIC) and the HIgh RESolution spectrograph (HIRES) started in 2016. ESO's contribution to their construction is in principle funded from the ELT operations budget line. The pivotal ExoPlanet Imaging Camera and Spectrograph (EPICS) requires extensive research and development (R&D), carried out in collaboration with institutes in the Member States.

2.2.3 Phase 2
Phase 2 for the ELT is currently unfunded but components can be proposed for Council approval if additional funding is identified. The prioritised list, established in mid-2014, of those components divides naturally into four groups:
i.  The LTAO module;
ii. The five inner rings of segments for M1 as well as the seventh sector needed for re-coating and efficient M1 maintenance;
iii. The second pre-focal station, needed for the fourth, fifth and sixth instruments, and beyond;
iv. Supporting systems, including power conditioning, additional buildings, the second M1 segment coating unit and equipment for astronomical site monitoring.

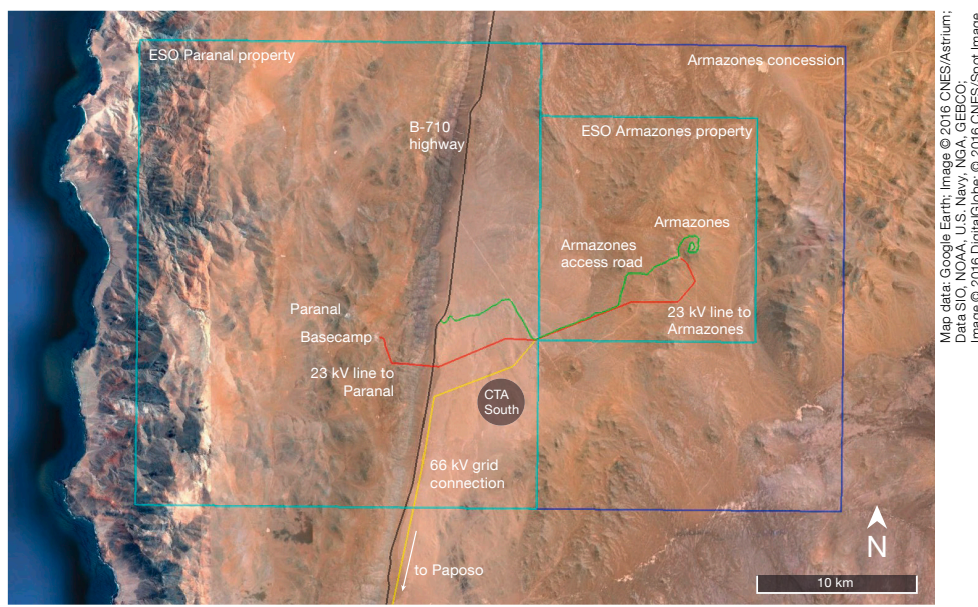

Figure 7. Map of the Paranal Armazones area. The new access road to Armazones is indicated (in green), together with the tracks of the grid connection from Paposo to the south, the distribution lines to Paranal and Armazones, and the location of CTA South.

The LTAO module is crucial for bringing HARMONI to its full potential. Adding the inner rings would increase the collecting area of M1 by 26 % and hence significantly increase the sensitivity of the ELT, thus decreasing the observing time required for a number of programmes while enabling others. The inner rings also impact the point spread function, with immediate consequences for instrument design and performance (for example, coronagraphy). It is therefore highly desirable to have all M1 segments in place before first light in 2024, and to have the additional seventh sector of segments in hand soon after. This also avoids interleaving (early) operations with additional construction activities, and simplifies the regular cleaning and coating of the M1 segments, thereby lowering costs. Adding the inner rings before first light also increases the value of the GTO nights for the instrument teams. The Phase 1 contracts related to M1 (blanks, polishing, segment support) therefore have an option for the provision of the missing segments and their support. The decision to exercise the option can be delayed to late 2019 without impacting the first light date, which provides time to find additional funding. Not exercising the option, and ordering the missing segments and their supports later through new contracts, would lead to a significant cost increase, and would heavily impact the initial science capabilities and operations of the ELT.

The Phase A studies of MOSAIC and HIRES will be completed in 2018 and by that time finalisation is required on the timing of their development and the way forward for the second pre-focal station which is needed for the deployment of both instruments.

The supporting systems, deferred to Phase 2, are needed to achieve the full performance of the ELT. These include: filtering out grid power cuts (needed from 2017); providing meteorological information to optimally schedule the observations and to assist in adaptive optics use cases; and reduction of the maintenance time to achieve the required mirror reflectivity. All these systems are aimed at minimising the technical downtime of the ELT.

2.2.4 Operations
The ELT will be operated as an integral part of the Paranal Observatory. The successful VLT end-to-end science operations model will continue to evolve to accommodate the needs of the community and the requirements of the ELT to ensure its efficient scientific exploitation. On-site science operations of the VLT and ELT will be fully integrated in the back- and front-end off-site support provided at Headquarters.

General operations support will be provided through the Paranal Director's Office and the Logistics and Facilities Department, with administrative support from the ESO Vitacura Office. This support includes site management, personnel, purchasing, finance, board and lodging, cleaning, transportation and commuting services, facilities, building and road maintenance, medical services, and the safety and security of the Observatory. Technical operations support is provided by the Maintenance, Support & Engineering Department and includes preventive and predictive maintenance of telescopes, instruments and supporting equipment, troubleshooting, corrective maintenance, refurbishments, provision of technical materials, consumables and spare parts, and project support. These tasks are carried out by a dedicated workforce, consisting of international staff and local staff members recruited in Chile. Many activities that are not considered ESO core competences are outsourced as service contracts to specialised companies. In addition, Headquarters provides personnel effort and support for regular and specialist maintenance and repair of telescopes, instruments and facilities.

The existing on-site operation capacity and capabilities, including the external service contracts, will be expanded and developed according to the projected needs of the ELT. While aiming at a fully integrated operation of the VLT and ELT to maximally exploit synergies, the accounting of operation cost and effort will remain separated for both programmes to prevent cross-subsidisation.

2.2.5 Paranal—Armazones infrastructure

In 2011, the original Paranal property donated to ESO by the Chilean government in 1995 was extended towards the east by the addition of a tract of land containing Cerro Armazones and a surrounding area given in concession for 50 years. Figure 7 provides a map of this area. A new access road connecting Armazones to the public road B-710 has been constructed, cutting the driving time from the Paranal base camp to the Armazones platform to about 40 minutes. The road is asphalted and 11 metres wide, allowing heavy construction trucks to pass each other. Part of the original access track up to the top of Armazones is now a service road running next to the high-voltage cable trench.

The construction and operation of the ELT require additional infrastructure, including technical buildings, storage areas and accommodation facilities. The telescope building on the Armazones platform will host only operationally critical facilities. The nearby Armazones base camp is dedicated to the needs of construction and will be dismantled when the ELT comes into operation. Therefore, most additional infrastructure will be implemented as extensions to the existing Paranal base camp. The majority of these new installations will already be required during the ELT assembly, integration and verification (AIV) phase, and will need to be maintained for ELT operations. After the end of AIV, the available temporary accommodation will allow VLT and ELT operation staff to be hosted, while the existing Residencia and the contractor's camp are being expanded and renewed to serve the increased need for accommodation, office space, meeting rooms and work places. This expansion would also take into account the needs of CTA South.

2.3 La Silla

The La Silla Observatory (Figure 8) successfully operates according to the streamlined and lean operations model endorsed by Council in June 2007. The La Silla 2010+ model supports the continued operations of the 3.6-metre and NTT telescopes, and their instrumentation, by ESO. Both telescopes continue to be highly oversubscribed.

The High Accuracy Radial velocity Planet Searcher (HARPS) on the 3.6-metre telescope leads the world in exoplanet hunting by means of radial velocity measurements and continues to be maintained and upgraded in collaboration with external institutes. The NTT operates with the ESO Faint Object Spectrograph and Camera 2 (EFOSC2) and the Son of ISAAC infrared camera (SOFI), and the telescope is still much in demand by the community. It is currently used for the PESSTO survey (Smartt et al., 2013) and also provides opportunities for novel instrumentation by offering a visitor focus.

Figure 8. A view of the La Silla Observatory.

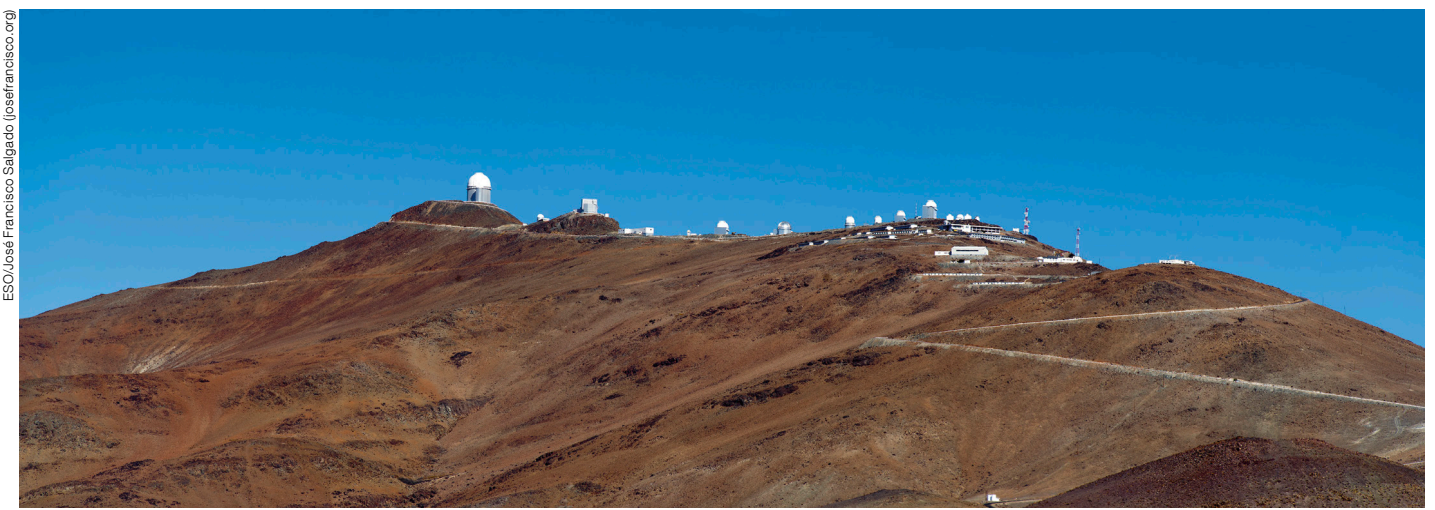

A call for new instruments was made in 2014, aimed primarily at replacing the ageing instrumentation at the NTT. The medium-resolution ($R = 5000$) optical and near-infrared (0.4–1.8 µm) spectrograph SOXS (Son of X-shooter) was selected as the future workhorse instrument at the NTT. SOXS addresses in particular — but not exclusively — the needs of the time-domain research community. Furthermore, the high-speed, triple-beam imager ULTRACAM, a visitor instrument, was offered for up to 25 % of NTT time in exchange for cash contributions to NTT operations. In addition, the Near Infra-Red Planet Searcher (NIRPS) was selected as the near-infrared extension of HARPS on the 3.6-metre telescope, creating the most powerful optical to near-infrared precision radial velocity machine for exoplanet research in the southern hemisphere.

An increasing number of small-telescope projects are hosted at La Silla. These scientific projects are developed by teams in the community; they are funded by national institutes or by the European Research Council or via private donations, and take advantage of the excellent atmospheric conditions, the available infrastructure and the lean operation-support model.

La Silla hosts the the Max Planck Gesellschaft MPG/ESO 2.2-metre telescope, the Danish 1.54-metre, the Swiss 1.2-metre Leonard Euler, the ESO 1-metre, the Rapid Eye Mount (REM), the Télescope à Action Rapide pour les Objets Transitoires (TAROT-S) and the TRAnsiting Planets and PlanetesImals Small Telescope (TRAPPIST), all with dedicated instruments. An increasing number of these telescopes are operated remotely. The QUEST survey project on the 1-metre ESO Schmidt telescope (Baltay et al., 2012) was completed after eight years and operation was discontinued as planned at the end of March 2016. Recently, the Universidad Católica Norte in collaboration with the Pontificia Universidad Católica de Chile has upgraded the ESO 1-metre telescope and installed the FIber Dual Echelle Optical Spectrograph (FIDEOS).

Two new projects, ExTrA (Exoplanets in Transit and their Atmospheres) and MASCARA (Multi-site All-Sky CAmeRA), are currently being added to the suite of hosted telescopes. All these projects can be kept cost-neutral to ESO through financial contributions by the project teams to the site operations costs, as long as the NTT and the 3.6-metre telescope are operated by ESO. An increasing fraction of the projects offer access to their high-level data products via the SAF in return for lowered hosting fees.

The La Silla 2010+ plan was envisaged for an initial period of five years without major re-investment into the site infrastructure. Modest re-investment is now needed to maintain the site infrastructure at the level required by the continued operation of the 3.6-metre, the NTT and the hosted telescopes. Operation-critical information technology (IT) infrastructure has been renewed during 2016. Savings have been made on electrical power costs, as La Silla became a regulated client in June 2015 in return for supporting the installation of a photovoltaic solar power plant on the premises. The plant started producing power in mid-2016 (Figure 9).

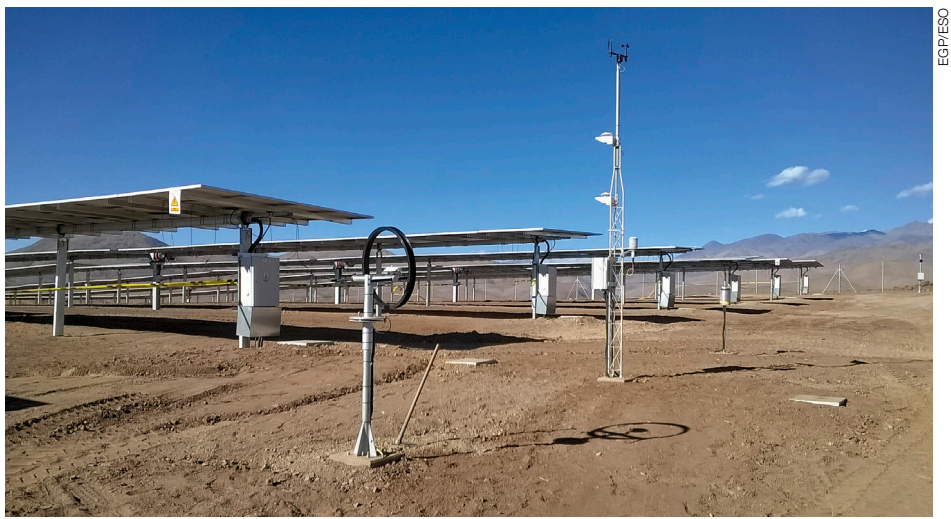

Figure 9. The new solar photovoltaic plant at La Silla.

The science operations model for La Silla — and the corresponding infrastructure support — are being evolved to maintain basic compatibility with its Paranal counterpart, so that the overall ESO end-to-end operations also encompass the ESO-operated facilities at La Silla. This support includes the tools and methods used by the community to prepare and execute observations at La Silla, or to exploit the data in the SAF.

The availability of SOXS on the NTT (and X-shooter on the VLT) will put the ESO community in an excellent position to follow up the most interesting transients to be discovered by the LSST from 2023 onwards. The combination of HARPS and NIRPS on the 3.6-metre telescope is crucial for providing critical ground-based complementary data for the ESA/Swiss mission CHaracterising ExOPlanet Satellite (CHEOPS) and for PLATO.

The extension of La Silla operations beyond 2020 as described above requires both NIRPS and SOXS to be successful. If NIRPS were to fail for some unforeseen reason, then the 3.6-metre telescope with HARPS would still be valuable for exoplanet research, but it would be reasonable for ESO to require external contributions to the operation costs. If SOXS were to fail, then the future of the NTT would be in serious doubt. This would threaten the viability of the entire La Silla operations model, as it is not cost-effective for ESO to run the complete site for a single medium-sized telescope. External funding or support could come from (consortia of) institutes in the Member States, or from partners elsewhere including the Host State Chile.

2.4 Technology development programme

It is critical to develop and secure key technologies which will maintain the Observatories at the cutting edge of astronomy and so contribute to achieving ESO's mission. In practice, this means taking those that are at low levels of technology readiness and developing them to a level sufficient to be incorporated within new projects with manageable risk. The requirement to not only develop but also secure technology for the future means that only rarely will a technology development project take the form of a conventional procurement from industry or institutes. Usually a collaborative approach is required to ensure that the intellectual property developed in the project is either transferred to ESO, or another scheme is used to ensure that the technology will be available and further developed over the period of time that it is needed. This approach requires long-term planning to address risks such as obsolescence, loss of external manufacturers, and uncertain requirements for future development.

In addition to enabling ESO's mission, technology development also has a role in enhancing the skills, and increasing the motivation, of staff within ESO. Although this aspect will not drive the selection of specific projects, it will be considered when deciding where and how the development will be carried out.

The funding for technology development has been consolidated from a variety of small R&D budgets across the Directorates of Programmes and Engineering, from the enabling technology budget for the VLT and from the enabling technology programme formerly planned for the ELT. Current projects include the development of deformable mirrors for future ESO projects, detector development, laser R&D, ELT real-time computing and wavefront-sensor camera development, and R&D for the ELT EPICS instrument.

The availability of state-of-the-art detectors is critical for ESO. Infrared devices have traditionally been single-source procurements from US companies who serve the defence market and are under significant ITAR (International Traffic in Arms Regulations) restrictions. The next-generation devices are extremely expensive and start to dominate the budgets and limit the scope of new instruments. In the optical, the situation is less urgent, but the traditional charge-coupled device (CCD) technology is now being replaced with Complementary Metal Oxide Semiconductor (CMOS) technology. This represents a viable alternative for commercial applications and in principle also for astronomy, but currently no such devices meet ESO's needs. A first step in mitigating these risks is participation in the ATTRACT initiative, which aims to obtain funding from the European Commission for developing imaging and detection technologies in Europe.

3. Submillimetre Programmes

ESO's activities in the submillimetre wavelength regime, traditionally associated with radio-astronomy, started with the Swedish-ESO Submillimetre Telescope (SEST) on La Silla (1987–2003). This led to participation in ALMA and APEX (Figure 10).

3.1 Chajnantor — ALMA

ALMA evolved from separate regional plans to a global partnership between ESO (37.5 %), the US National Science Foundation (NSF, representing USA, Canada and Taiwan; 37.5 %) and the National Institutes of Natural Sciences in Japan (NINS, representing Japan, South Korea and Taiwan; 25 %). The host state Chile receives 10 % of the observing time. ESO's counterparts at the executive level are the US National Radio Astronomy Observatory (NRAO), managed by the Association of Universities Inc. (AUI), and the National Astronomical Observatory of Japan (NAOJ).

3.1.1 Construction
ALMA construction on the Array Operations Site (AOS) at 5050 metres altitude and the Operations Support Facility (OSF) at 2950 metres formally ended in December 2013, except for, as regards ESO, the ALMA Residence and a site-security surveillance system. NAOJ completed delivery of the Band 4, 8 and 10 receivers soon after this date. NRAO continued after 2013 with the investigation of the astigmatism affecting the North American antenna performance and a number of other activities. Commissioning activities of the array are continuing and extend into the operations phase for more advanced observing modes, including solar observations and participation in global very long baseline interferometry (VLBI) experiments.

The ALMA Residence will be completed and handed over to ALMA in early 2017, and procurement of the site-security surveillance system will follow. The permanent power system started operations in November 2012 and is now in reliable 24/7 operation, after some initial technical and operational difficulties in 2013. Previous ESO infrastructure deliveries include the Technical Building for the OSF, the Santiago Central Office (SCO) at the ESO Vitacura premises, 192 antenna foundations, and the access road to the OSF and AOS.

Front End and Back End deliveries were complete by the end of 2013. These included 73 receiver cartridges each for Band 7 and Band 9, 58 water vapour radiometers, 83 Front End power supplies, 26 fully integrated and tested Front End assemblies, 70 cryostats, more than 500 cartridge bodies, more than 600 photo mixers, the complete fibre management system, 550 tuneable filter boards, and 58 sets of cryogenic helium lines. Two custom-made antenna transporters have been in routine operation since 2008. The contract with the AEM Consortium for the antennas provided by ESO was formally closed in early 2016, when the warranty period of the twenty-fifth antenna expired. All AEM antennas are within specification and operating reliably.

Council set the cost-to-completion for ESO's ALMA construction in 2005, at an amount equal to 489.5 MEUR in 2014 prices. The entire programme was delivered over the next nine years with only a 1.5 % cost increase. A workshop on

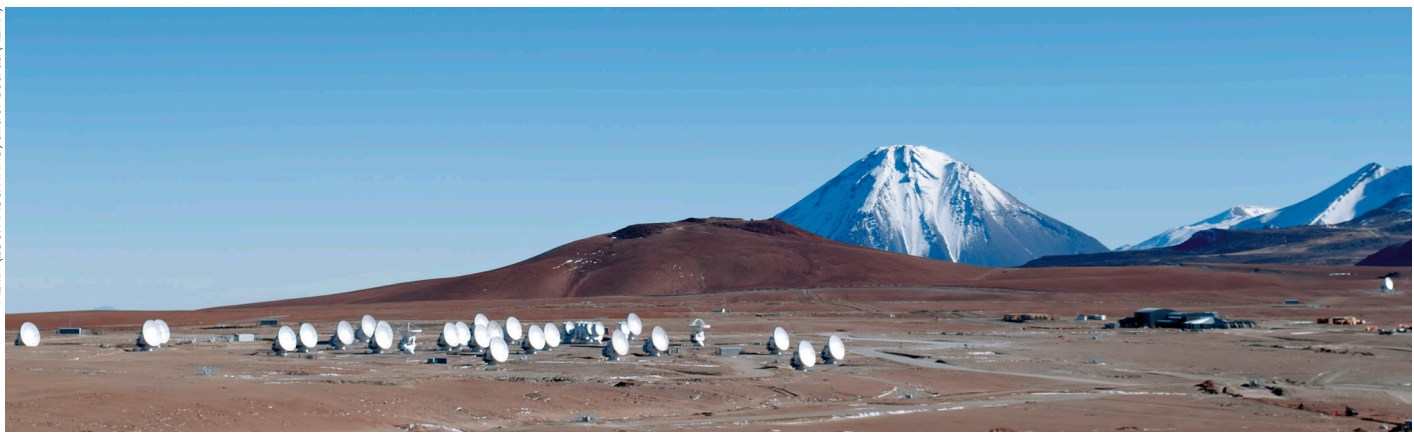

Figure 10. The ALMA antennas on the Chajnantor plateau, with APEX visible on the right. The Licancabur volcano towers in the background.

Lessons Learned from ALMA Construction[5] took place in 2015, and the results are being applied in the ELT programme.

The AEM prototype antenna, which was tested at the NRAO site near Socorro, New Mexico in the period 2006–2009, was transferred by ESO to Steward Observatory in Tucson, to be part of the Arizona Radio Observatory. It has replaced the venerable 12-metre antenna on Kitt Peak. In return, the ESO community receives a total of 3600 hours of guaranteed time on the refurbished prototype and on the 10-metre Heinrich Hertz Submillimeter Telescope on Mount Graham.

3.1.2 Operations
ALMA on-site operations are carried out by the Joint ALMA Observatory (JAO), whose personnel comprises both international and local staff, and are complemented by off-site operations activities at the three executives. The scientific results are transformational but there is still further to go to reach ALMA's full scientific potential and activities continue to test advanced modes of observation. Improvements in observing and data reduction efficiency are needed to achieve a long-term sustainable operations model.

The ALMA-wide governance for the operations phase has been agreed and established. In December 2015, the ALMA Trilateral Agreement was signed by ESO, the US National Science Foundation (NSF) and the Japanese National Institutes of Natural Sciences (NINS). It supersedes previous bilateral agreements between the ALMA partners. For the daily interactions and management of ALMA, a structure of Integrated Teams across all ALMA partners and JAO has been established. These Integrated Teams operate for Management, Science, Science Operations, Engineering, Computing, and Outreach, replacing the construction-oriented Integrated Product Teams (IPTs).

ALMA operations support at ESO formally started in early 2008 with the establishment of the European ALMA Regional Centre (ARC). The ESO ALMA Support Centre (EASC) was created in 2009 to centrally coordinate the overall ESO off-site operation support. It provides user support, data processing, technical hardware and software support and maintenance, a development study programme and upgrades for ALMA, and the delivery and archiving of data. The EASC also supports ALMA Science and Outreach activities.

ALMA Early Science observations started with Cycle 0 in September 2011 using 16 antennas, leading to a number of high-profile discoveries including the detection of interstellar sugar, planet-forming discs, and water in star-forming galaxies at redshift 5.7. During the long-baseline campaign in late 2014, an iconic image was obtained which reveals unprecedented fine detail in a planet-forming disc around the young star HL Tauri (see Figure, p.27). It attracted tremendous public interest worldwide.

ALMA science operations adopted an annual cycle starting with Cycle 3 in October 2015. Cycle 4 has configurations comprising at least forty 12-metre antennas, as well as the 7-metre antennas of the Morita Array (previously known as the Atacama Compact Array), extended baselines and seven (of the final ten) receiver bands. The oversubscription continues to be highest for the ESO Member State astronomers and they have had involvement in around three quarters of the ALMA papers already published. The ARC is engaged in helping ALMA users to prepare their observations and reduce the resulting data, and in testing of observation preparation and data processing software.

The distributed model for the ESO ARC, with the central hub at ESO and a network of locally funded nodes in several Member States, has proven to work very well (Hatzminaoglou et al., 2015). It provides ALMA users with high-quality data products in a timely manner, in addition to face-to-face support. An external review in January 2015 confirmed the value of this model while also outlining some risks and challenges for the future. One of these is the potential closing of an ARC node and the corresponding reduction in data processing effort.

In the area of data reduction, the calibration pipeline has been in use since late 2014, but, contrary to initial plans, the imaging pipeline is not yet available. This poses a significant challenge for the ARCs. It is essential that more than 75 % of the data can be efficiently pipeline processed in the near future to ensure that the workload of the ARC staff remains manageable and they can continue to provide a high level of support to users.

ESO has undertaken (with support of the ALMA partners) the planning for, and construction of, a pipeline to connect the OSF turbines to the nearby North Andino gas pipeline. This will provide natural gas at a competitive price, reducing operational costs. It will also reduce the traffic of heavy trucks on the road to the OSF (which lowers maintenance needs and increases safety), improves the working efficiency of the multi-fuel turbines and limits pollution.

3.1.3 Further developments

The current ALMA operations plan contains off-site development funding of which ESO's 37.5 % share ramped up to approximately 3.2 MEUR/year in 2015. The use of this budget for development efforts in the software and hardware areas follows the agreed principles for the ALMA Development Programme, which include an ALMA-wide approval process and governance. The EASC plays a key role in representing the interests of the ESO Member State communities and managing the ESO part of the development programme.

ESO issued calls for ALMA development studies to the technical and scientific community in the Member States in 2010, 2013 and 2016. These studies include science investigations (such as the scientific value of a possible Band 11), design studies for new hardware and software, and preparations for series production of components. Similar initiatives were taken in North America and East Asia. The first resulting full-scale projects are underway or already completed (for example, the fibre link from the OSF to SCO). ESO has taken the lead on the Band 5 project following the successful development and pre-production of six receivers funded over the period 2008 to 2012 by the European Union Sixth Framework Programme for the ALMA Enhancement. Within the ALMA Development Programme, ESO provides 73 Band 5 cartridges and NRAO provides the local oscillator (as in the ALMA construction phase). The Band 5 cartridge work is contracted to a European consortium and started in 2012. Equipping all 66 ALMA antennas with Band 5 receivers should be finished in late 2017. The integrated alarm system to be developed by ESO will be operational by the end of 2018, improving the overall reliability and safety of the Observatory.

In the coming decade, ALMA will begin upgrading the receivers and signal transport and processing chain. ESO's ALMA development studies are already addressing the possibilities of building a next-generation receiver for the 67–116 GHz range (Band 2+3), as well as upgrades to Bands 7 and 9. The development of a new digitiser system is also well advanced. Studies of possible upgrades to the software systems are also being carried out.

ALMA-wide activities are under way to define a coordinated long-term development plan across the ALMA Partnership to ensure maximum gain for the available funding. The ALMA Scientific Advisory Committee (ASAC) and ALMA programme scientists have elaborated development paths in the ALMA 2030 process resulting in three documents, covering: 1) the major science themes in the 2020–2030 decade; 2) the landscape of major facilities by 2030; and 3) the pathways to developing ALMA. In 2015, the ALMA Board created an ALMA Development Vision Working Group to translate these plans into a realistic path for ALMA development. A draft report is expected in April 2017.

On ESO's side no major upgrade of ALMA (other than those funded in the development plan) is foreseen until the ELT construction is completed, unless an additional partner joins ALMA. It is noted that this would further increase the oversubscription for observing time.

3.2 Chajnantor — APEX

The Atacama Pathfinder Experiment (APEX) is a 12-metre antenna for submillimetre astronomy located at an altitude of 5065 metres on Llano Chajnantor in northern Chile. APEX is a partnership between the Max-Planck-Gesellschaft (MPG), the Onsala Space Observatory (OSO) and ESO, with 50, 23 and 27 % shares respectively, and it saw first light in 2006. It is operated by ESO from a base in Sequitor near San Pedro de Atacama, as part of the La Silla Paranal Observatory.

APEX celebrated its tenth year of science operations in January 2016. It is an ideal wide-field mapping facility, complementary to ALMA, and allows testing of innovative instrumentation. It has an important role as source finder for ALMA and also enables fast spectroscopic follow-up of ALMA discoveries. APEX continues to be highly oversubscribed and produces a steady output of scientific publications. APEX is operated in service mode and employs support astronomers from the partners and Chile; it has consequently become an important facility to train a new generation of astronomers in submillimetre-wavelength expertise.

The APEX Partnership is currently agreed until the end of 2017. The partners have reviewed the possibility of extending the agreement for five years from 2018 to 2022. An external critical review was carried out in early 2016 to assess the scientific and technical competitiveness of APEX beyond 2017. With the positive feedback from the review, the ESO Council has just approved a new agreement for a further extension through to 2022. As part of this agreement, the APEX Partnership plans for additional investments in antennas, instrumentation and infrastructure to ensure the scientific and technical competitiveness of the facility. The aim is to offer a state-of-the-art suite of heterodyne receivers across the accessible submillimetre window, together with kilopixel continuum multi-colour widefield cameras. The partnership will adjust the partner shares of MPG/ESO/OSO to 55 % / 32 % / 13 % as of 2018. The cost to ESO of the APEX extension is included in the financial planning (see Figure 16).

If the APEX Partnership decided to cease operations after 2022, the facility could either be decommissioned and removed from the site or could transition to a new initiative that builds on the existing APEX infrastructure at Chajnantor and Sequitor and on the available expertise in operating such a facility. A large single-dish submillimetre antenna with massive widefield continuum imaging capabilities could be such a future initiative.

### 4. Participation in a high-energy programme

Very high energy gamma radiation from celestial objects can be studied from the ground by observing the optical Cherenkov flashes generated when the gamma rays interact with the Earth's atmosphere. Pioneering work in this area was done by the High Energy Stereoscopic System (HESS), the Major Atmospheric Gamma Imaging Cherenkov (MAGIC) and the Very Energetic Radiation Imaging Telescope Array System (VERITAS) experiments. The CTA project aims to take the next step in energy range, sensitivity and resolution by means of a northern and a southern array of many simple, but large, open-air optical telescopes with segmented mirrors.

### 4.1 CTA project

The scientific goals of CTA are extremely broad, ranging from understanding the role of relativistic cosmic particles to the search for dark matter and probing environments from the immediate neighbourhood of black holes to the cosmic voids on the largest scales. Covering a huge range in photon energy from 20 GeV to 300 TeV, CTA will improve on all aspects of performance with respect to the precursor experiments. Wider field-of-view and improved sensitivity will enable CTA to survey hundreds of times faster than previous TeV instruments. The angular resolution of CTA will approach one arcminute at high energies — the best resolution of any space or ground instrument operating above the X-ray band — allowing detailed imaging of a large number of gamma-ray sources. An improvement in collecting area of between one and two orders of magnitude also makes CTA a powerful instrument for time-domain astrophysics, and three orders of magnitude more sensitive on hourly timescales than the Large Area Telescope on the Fermi Gamma-ray Space Telescope at 30 GeV. The two arrays will provide full sky coverage, and hence maximise the potential for observing rare phenomena including the nearest supernovae, gamma-ray bursts or gravitational wave transients.

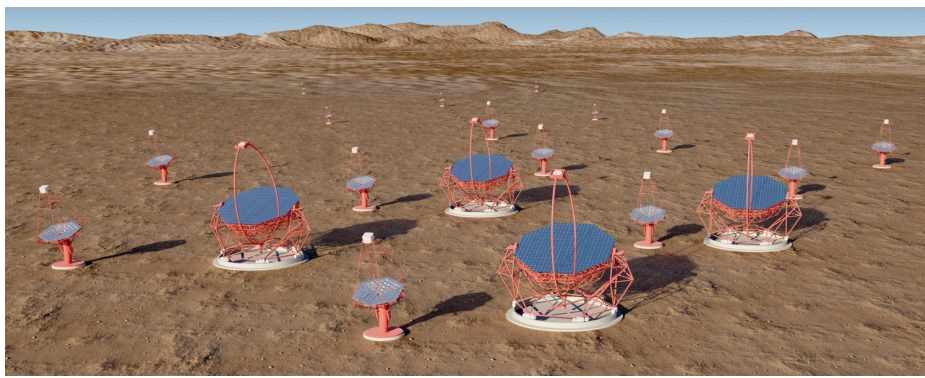

Figure 11. Artist's impression of the Cherenkov Telescope Array South at Paranal.

The CTA project carried out a worldwide site selection process in 2013–2014 which resulted in the identification of La Palma as the preferred northern site and the Paranal area as the ideal southern site (the location is shown in Figure 7). In the baseline plan, the southern array will cover approximately ten square kilometres, and will consist of telescopes of three sizes: 70 small (4-metre), 25 medium (12-metre) and 4 large (23-metre). The northern array will have 15 medium-sized and 4 large telescopes. Construction is expected to take approximately four years.

Unlike the precursor experiments, both arrays will be operated as a proposal-driven facility, with all data available in a public archive after a one-year proprietary period. A core programme of key science projects will use about 40 % of the available observing time in the initial years to provide legacy data products. With 99 telescopes on the southern site, very flexible operation will be possible, with sub-arrays available for specific tasks.

The CTA Partnership consists of a Governing Council, a Consortium and a Project Office. The CTA Consortium comprises over 1200 scientists working in 200 institutes from 32 countries, among which are 12 ESO Member States as well as Brazil and Chile. In mid-2016, the Istituto Nazionale di Astrofisica (INAF) centre in Bologna was selected as the site for the CTAO Headquarters and the Deutsches Elektronen-SYnchrotron (DESY) centre at Zeuthen (Germany) as the site for the Science Data Management Centre. The Partnership is making progress in securing funding for construction, and expects to start with partial arrays in each location. These will already be able to do exciting science by increasing the exposure time.

### 4.2 Role for ESO

The ESO financial plan does not include funding for construction and operation of new facilities on the timescale planned for CTA. However, ESO has tremendous experience in maintaining and operating optical telescopes in remote areas, which is valuable for the CTA project, and will enhance the science return on the investment in CTA.

Following the identification of the Paranal area as the preferred southern site, detailed discussions took place with the CTA Partnership to clarify the practicalities of hosting the southern array on the Paranal property, and to develop a formal agreement between ESO and CTA. The agreement stipulates that ESO joins the CTA Partnership with an 8 % voting share, and operates the southern array on behalf of the partnership on a cost-neutral basis, with funding provided by the CTA Partners. In return, 10 % of the observing time on both the southern and the northern array on La Palma would be made available to scientists in the ESO Member States (in addition to the CTA partner shares that 12 of the 15 Member States have), and 10 % of the observing time on the southern array would be reserved for Chilean scientists. The agreement was ratified by the ESO Council in its December 2016 meeting. If ESO were to further cover the cost of site operations (estimated at about 4 MEUR per year), this could add approximately 10% of observing time to the ESO share, but no such funding is identified at present.

The operation and technical facilities of CTA will be located close to the telescope array itself unless clear synergies with ELT and VLT facilities at the Paranal base camp can be identified. Provision of accommodation to CTA staff in the expanded Paranal facilities is one of the most obvious possibilities.

Operating CTA South provides an exciting expansion of the overall programme, opening a new window on the Universe for astronomers in the Member States. CTA South will be integrated into the Paranal Observatory, which will guarantee its sustainability and control any risks of conflict for ESO resources during ELT construction. It is also fully in line with ESO's mission of building and operating world-class facilities for astronomy and fostering collaboration in astronomy.

## 5. The Organisation

Since its foundation in 1962, ESO has established the structure required to conceive, develop, build and operate advanced ground-based astronomical observatories. It has created a close collaboration with the Member States in respect of the supporting astronomical and technical research and development work, modelling, systems engineering, etc. necessary to deliver not only the hardware but also software and pipeline-reduced data to its community. The 15 Member States with Brazil and Chile account for about one third of the world's astronomical community.

### 5.1 Organisational structure

ESO Headquarters is the focus for interaction between the Member States and the Organisation and provides support for the governing and advisory bodies. It hosts approximately 420 personnel and is the centre for development of telescopes and instruments, operations support, archives, scientific and technical research, education and public outreach, human resources, financial management and procurement. It includes data centres and laboratories for the development of key technologies as well as state-of-the-art integration halls and equipment to test instruments before shipping to the Observatories. A map of the Headquarters buildings is shown in Figure 12.

Approximately 275 personnel work in Chile, distributed over the three sites of the La Silla Paranal Observatory (La Silla, Paranal and Sequitor), the ALMA sites, and the ESO premises in Vitacura (Figure 14). The latter covers administrative activities, human resources, public outreach and official representation in Chile, together with the ALMA Santiago Central Office, and provides an environment for the fostering of science for staff astronomers and visitors. The ESO Guesthouse hosts visitors and Garching staff on their way to and from the sites and provides an additional channel for interaction with the community (Figure 15).

ESO's organisational structure consists of five Directorates: Science, Operations, Programmes, Engineering, and Administration. The Science Directorate provides scientific oversight across all programmes in the Organisation, handles observing proposal selection and telescope time allocation, education and public outreach and provides a research environment for interaction with the community. The Operations Directorate handles all La Silla Paranal Observatory needs, from user support for execution of science observations to delivery of pipeline-reduced science data, as well as the ESO support of ALMA operations. The Programmes Directorate contains programme and project management to define and execute major telescope and instrument projects in collaboration with the communities and industries in the Member States. The Engineering Directorate provides resources and services for the design, manufacturing, installation, corrective maintenance, upgrades and support of the telescopes, instruments and utility systems, and provides IT services across the Organisation. The Directorate of Administration includes human resources, finance, contracts & procurements and facility, logistics & transport. The Cabinet of the Director General, the Internal Audit Office and the Office for Representation in Chile provide support for the ESO Council and Finance Committee and to the Director General. A more detailed description can be found in the ESO High Level Organisational Structure document[6].

Staff astronomers at ESO are required to maintain an active research profile to be able to fulfil their functional duties. The studentship, fellowship, visitor and workshop programmes provide opportunities for close interaction with the astronomical community. The fellowship programme has been in place for 40 years and has trained many young scientists who return to institutions in the Member States. The support of VLT and ALMA operations by ESO fellows provides them with valuable experience for their careers. A pilot engineering fellowship scheme will start in 2017 and is expected to similarly increase the interaction with the engineering community in the Member States.

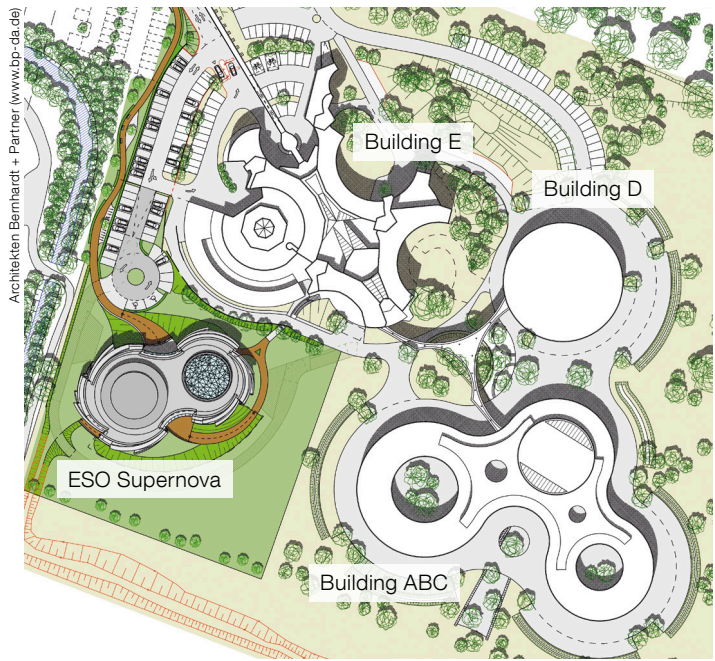

Figure 12. Map of ESO Headquarters in Garching. Building E is the original Headquarters building, connected by bridge to the extension Buildings ABC (containing offices, a larger auditorium, and a new Council Room) and D (the Technical Building, containing a new, larger integration hall).

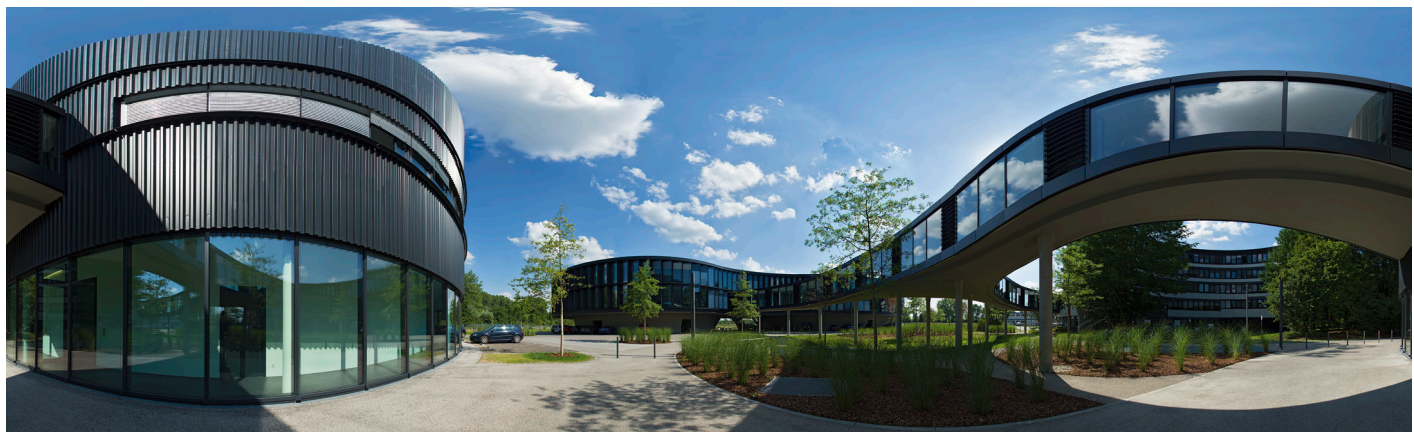

Figure 13. A 360-degree panorama of the ESO Headquarters, Garching, Germany.

Much work was done in the past decade to define and further develop internal processes and policies. These include, for example, the ESO People Policy, the ESO Code of Conduct, new Financial Rules and Regulations, Rules of Procedure for all governing bodies, a Technology Transfer policy, a Data Classification policy, a new ESO Safety Policy with associated safety manuals for Headquarters and the Observatory sites, translations of the Basic Texts into the languages of all the Member States, and a Protocol Manual. To optimise the structure of the Organisation in order to efficiently develop concurrent projects in a resource-constrained framework, the matrixing of engineering services was implemented. This development led to clarification of the role of programme and project management, one of the results of which was the ESO Project Management Framework, defining how ESO carries out projects in the era of ELT construction.

The Representation in Chile represents the Director General in all matters with the Chilean governmental, regional and local authorities, as well as with the diplomatic legations of the Member States in Chile. A newsletter for foreign missions in Chile keeps the legations up-to-date with scientific, strategic and organisational developments and builds knowledge about ESO.

The Cabinet of the Director General works on a broad range of projects including legal and international affairs, internal communication, corporate risks and intellectual property management, and protocol. The Cabinet also produces the Annual Report, the quarterly journal The Messenger, and the ESO Science Newsletter. Established over 40 years ago, The Messenger continues to provide news of ESO's telescopes and instrumentation, a selection of recent scientific work carried out with ESO facilities, and reports on workshops, along with profiles of ESO fellows. More rapid communications on all scientific activities are provided by the Science Newsletter[7], which collects announcements on the science webpage into a newsletter mailed approximately monthly to about 9000 subscribers with ESO User Portal accounts. The internal communication office supports and coordinates communication flow across ESO, including ESO-wide internal announcements.

5.2 Evolution of staff skills

In the past three years the core competences required to carry out ESO's programme were identified, looking forward 15 years and involving extensive consultation with the community. These core competences depend on the top-level requirements of the ESO projects, on the strategic share of work between ESO and industry or technical and scientific institutions in the Member States, and on technical developments worldwide. ESO's overall core competence is the ability to design, build and operate state-of-the-art astronomical facilities that produce front-line science. As a large part of the work is done by the community and by industry, the ESO scientists and engineers must have the system knowledge and the experience required to initiate, and effectively manage, the procurement of telescopes, instruments and associated equipment. Engineers must be capable of carrying out specialised design and analysis in telescopes and instrumentation, while scientists must be available to provide cradle-to-grave operational support to the Observatories. Finally, the community relies on ESO to procure specific items such as detector systems.

Developing and maintaining these core skills requires training as well as increased internal mobility. This has the added advantage of further increasing the motivation and drive of the staff. Areas of attention include adaptive optics technologies, real-time computing, laser development, wavefront control, phasing & metrology, ultra-high-contrast imaging, as well as expertise in project management and instrument systems engineering.

ESO will continue to participate in the development of instrumentation and other facilities, delivering major subsystems as a consortium partner. The goal is to lead at least one project at any time. This aim depends on available resources, but is key to the long-term retention and development of ESO staff, and the ability to work effectively with the community.

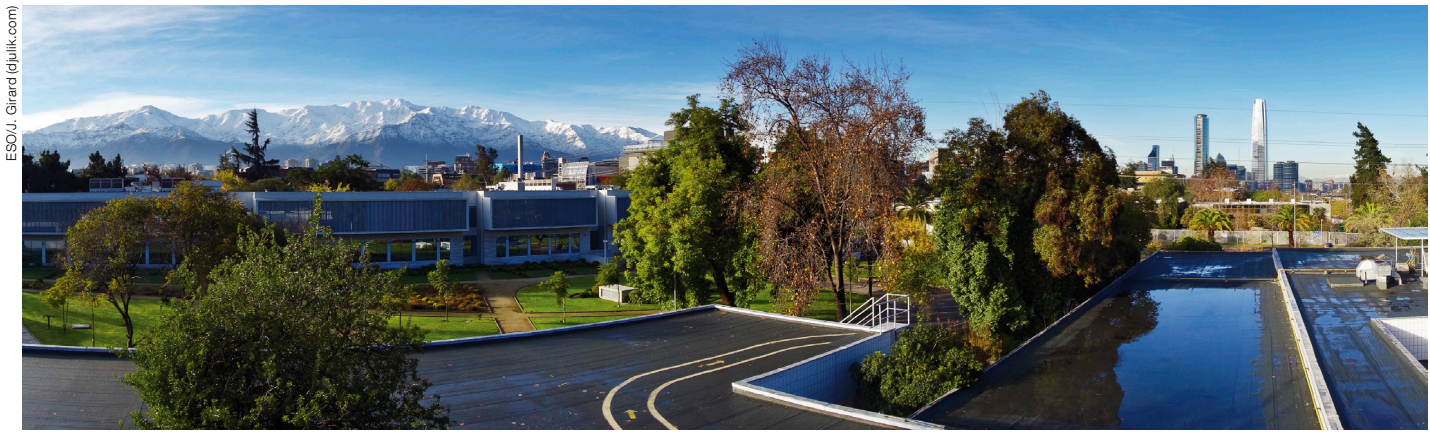

Figure 14. The Santiago site, located in Vitacura, houses the ESO offices and the ALMA Santiago Central Office.

ESO built the VLT test camera, the Infrared Spectrometer And Array Camera (ISAAC), the UV-Visual Echelle Spectrograph (UVES) and HAWK-I for the VLT in-house, and was a consortium partner in X-shooter. The Organisation has been a partner on every instrument built for the VLT, delivering the detector systems and often also key cryo-vacuum and AO subsystems. ESO staff have led the AOF project for UT4 from its start in 2004, including the development of the 4LGSF, and carried out the VLTI facility upgrade project in 2015 to prepare for the arrival of GRAVITY and MATISSE. Once these are completed, the test camera for the ELT will be built in-house and will require the full suite of core competences. Furthermore, the technology development programme, as well as prototyping activities, will provide ESO staff with additional means to develop their hands-on experience. For instance, it is natural for ESO to coordinate the technology development needed to bring the EPICS concept, with its technology readiness, to the level needed to start construction in 2020.

### 5.3 Age distribution and gender balance

A long period of sustained increases in staff complement ended in 2011, when ALMA construction activities ramped down at a time when ELT construction had not yet started (see Figure 17). This meant a reduced influx of junior staff, with, as a net result, an increase in the median age of ESO's international staff and a significant increase in the ratio of indefinite to fixed-term contracts. This will need attention in the years to come, as it is important to ensure an appropriate mix of junior and senior staff and to have the ability to react in a timely fashion to changes in qualifications and skills required by operations and projects, taking into account the required evolution of the skill mix previously described.

Gender balance is good in the administration area. It is also good amongst the population of students and fellows, but decreases amongst the more senior science and Observatory support staff. It is poor in the engineering and project management areas, and in senior management. This depends on a number of factors, some of which are beyond ESO's control. Work continues on improving gender balance, in particular within the technical and scientific professions as well as in the management of the Organisation. A gender diversity and inclusion plan is being prepared around three pillars, namely the recruitment and career development process, an enabling environment for women, and an improvement in the working conditions. A gender steering committee has been created and the existing participation in gender networks will be intensified.

### 5.4 Collaborations

ESO works closely with its communities and has a number of collaborations and partnerships, including those for instrumentation development and for ALMA support, as well as for cross-allocation of observing time with ESA, education, outreach and other activities.

ESO is a member of EIROforum, a partnership of eight of Europe's premier intergovernmental scientific organisations (the European Organisation for Nuclear Research [CERN], the European Molecular Biology Laboratory [EMBL], ESA, ESO, the European Synchrotron Radiation Facility [ESRF], the European Consortium for the Development of Fusion Energy [EUROfusion], Institut Laue–Langevin [ILL] and the European XFEL Free-Electron Laser Facility). These organisations have similar governance structures and serve a huge scientific community. As EIROforum, they support European science by sharing their experience, resources and facilities. Together, they interact with the European Commission, national governments, industry, science teachers, students and journalists. ESO chairs EIROforum from 1 July 2016 to 30 June 2017.

ESO has a considerable overlap of interests with ESA, Europe's leader in space research and technology, and with CERN, the pre-eminent centre for particle physics research. CERN and ESO share common interests in many technologies, and the potential for fruitful scientific collaboration is growing as the astroparticle and fundamental physics communities draw closer to astronomy. Joint ESO–ESA activities have been in place for a long time, including cross-allocation of observing time between the XMM/Newton X-ray observatory and the VLT, nightly VST monitoring of the position of the Gaia spacecraft, as well as ambitious public spectroscopic surveys such as Gaia-ESO, which have enhanced the tremendous impact of both Gaia and VLT.

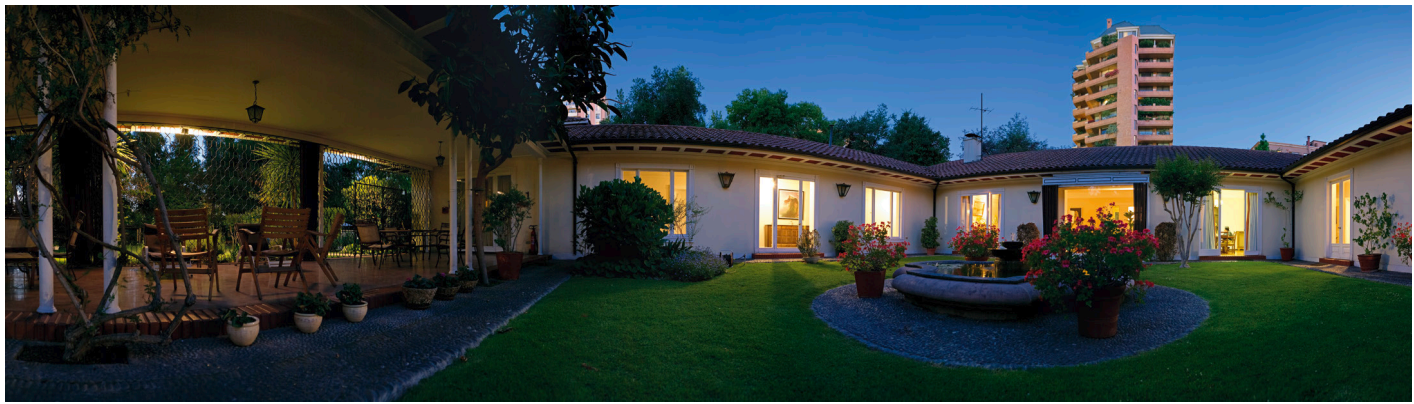

Figure 15. The garden of the Santiago Guesthouse.

From 1981 to 2010, ESO hosted the Space Telescope European Coordinating Facility, with which the ESO Science Archive Facility was jointly developed, and ESO continues to generate outreach material in support of the Hubble Space Telescope. The next generation of L(arge)- and M(edium)-class ESA missions will require even closer cooperation, since provision of supporting ground-based data is evolving from desirable to essential. An example is the crucial role of ground-based radial velocities for the transiting exoplanets that will be found by PLATO. The organisations also share a number of technology development goals, such as curved optical detectors, and high-speed, low-noise infrared detectors. ESO concluded two cooperation agreements, one with ESA in August 2015, which includes the exchange of observers for the relevant advisory bodies, and another one with CERN in December 2015, providing a framework for closer bilateral cooperation and exchange of information. ESA and CERN had previously concluded a bilateral agreement in March 2014.

The ESO–ESA and ESO–CERN agreements address scientific research, technology, and education and public outreach activities and promote the strategic coordination of the three organisations' long-term plans, as well as the coordination of scientific and training programmes. Both agreements encourage the coordination of services, tools and resources, in addition to the sharing of best practice in many areas. The organisation of joint seminars and workshops is another area of coordination, along with possible exchange of staff.

The Astronomy & Astrophysics (A&A) Journal is one of the leading astronomy publications. It was created in 1969 by combining a number of national journals. From the start, ESO has acted on behalf of the A&A Board of Directors in contractual matters and provides legal and administrative services, with the restriction that ESO does not commit itself to any direct financial sponsorship of the Journal, nor does it interfere in the scientific policy of the Journal. In return, ESO has one representative on the A&A Board of Directors. In view of the accession of numerous sponsoring bodies to the Journal, a renewal of the initial agreement between ESO and the sponsoring bodies became necessary in order to establish new terms and conditions for the continued publication of the Journal; the new agreement entered into force on 1 March 2016.

ESO collaborates with the International Astronomical Union (IAU) on topics related to outreach and supports the IAU web site. ESO is an observer on the United Nations Committee for the Peaceful Uses of Outer Space and is a member of the International Asteroid Warning Network which coordinates monitoring of potentially hazardous near-Earth objects.

ESO is also involved in various activities that aim to coordinate long-term planning of European astronomy and astroparticle physics through participation in the Optical Infrared Co-ordination Network for astronomy (OPTICON), RadioNet, ASTRONET and the Astroparticle Physics European Consortium (APPEC) networks. Several ESO staff hold European Research Council grants.

5.5 Outreach

ESO plays an important role in stimulating astronomical awareness through its education and outreach programme, which is coordinated closely with similar activities in the Member States. Outreach to the media and general public increases the visibility of, and support for, ESO and astronomy in the Member States and beyond. Harnessing the enduring appeal of astronomical discoveries can also draw young people into science and technology and contribute further to the efforts of the Member States in developing a technologically literate workforce, capable of meeting their future skill needs in a high-tech environment.

Public outreach tuned to the interests of the Chilean public and authorities is an essential component in the relationship between ESO and its Host State, giving visibility to the benefits that Chile's partnership with ESO has brought to the country's scientific development. It is also necessary to maintain ESO's identity among the major projects (for example, LSST and GMT) being built in Chile by other organisations.

The outreach activities will acquire an additional dimension in 2018 when the ESO Supernova Planetarium & Visitor Centre — a cutting-edge astronomy centre for the public, with free access — opens its doors. The heart of the ESO Supernova is a planetarium with state-of-the-art projection technology and a scientifically accurate three-dimensional astronomical database, ensuring a unique and authentic immersive experience. A large exhibition space and seminar rooms for interactive workshops provide the opportunity for ESO to become a leader in supporting and enhancing science education and literacy. Training workshops at the ESO Supernova Planetarium & Visitor Centre will inspire, and engage directly with, primary and secondary teachers from the Member States. ESO will provide key resources, encouraging teachers to interact with ESO's scientists and engineers, thereby fully exploiting the educational multiplier where teachers bring back their enthusiasm and knowledge to the classroom and thence to the tens to hundreds of pupils they teach each year.

The number of public weekend visitors to Paranal is approximately 8000 per year, and is expected to grow further when ELT construction activities ramp up on Armazones (and for CTA South). The current visitor centre is inadequate for this influx, and a new centre has been designed. There is no provision for construction funds in the long-term financial plan, but the hope is that external funding can be raised to realise this visitor centre in the near future. Activities are in train aimed at interesting (Chilean) philanthropists in this opportunity. Some of the content developed for the ESO Supernova Planetarium & Visitor Centre could be re-used here.

In the late 1990s, ESO and the municipality of Vitacura agreed to build an astronomy museum in the future Parque Bicentenario, close to the ESO premises. The park was established in 2007–11 and has become a popular recreational area for eastern Santiago. External funding to build and operate the museum would help ESO to honour its commitment and would provide the opportunity to promote astronomy and ESO at this optimal location.

## 6. Financial planning

The baseline plan for the next 15 years assumes that ESO's income is provided by the 15 Member States, contributing according to the principles for funding the ELT programme approved by Council: a year-on-year 2 % increase in the contributions by the Member States, on top of normal indexation, starting in 2013 and continuing through 2022; an additional contribution of 268.5 MEUR (2017 prices) by the Member States in the period 2013–2022, spread over the ten-year period. This baseline plan does not include any contribution from Brazil or from new Member States.

The planning assumes the following boundary conditions on the overall programme:
– Only the construction of the Phase 1 ELT and its instruments is included;
– Paranal operations and instrumentation are fully protected;
– The ALMA contribution remains at the 2016 level (including development);
– La Silla income includes a stable contribution from partners for hosted telescopes;
– APEX participation continues through 2022;
– Participation in CTA is cost-neutral for ESO.

In addition, the following assumptions are made:
– The cost of the Phase 1 ELT, including contingency, is 1033 MEUR (2016 prices) with first light in 2024;
– The special organisational contribution to the CERN Pension Fund is set at 1.3 MCHF per annum for the entire period considered;

Figure 16. Expenditure and overall income for the period 2016–2040 in graphical form.

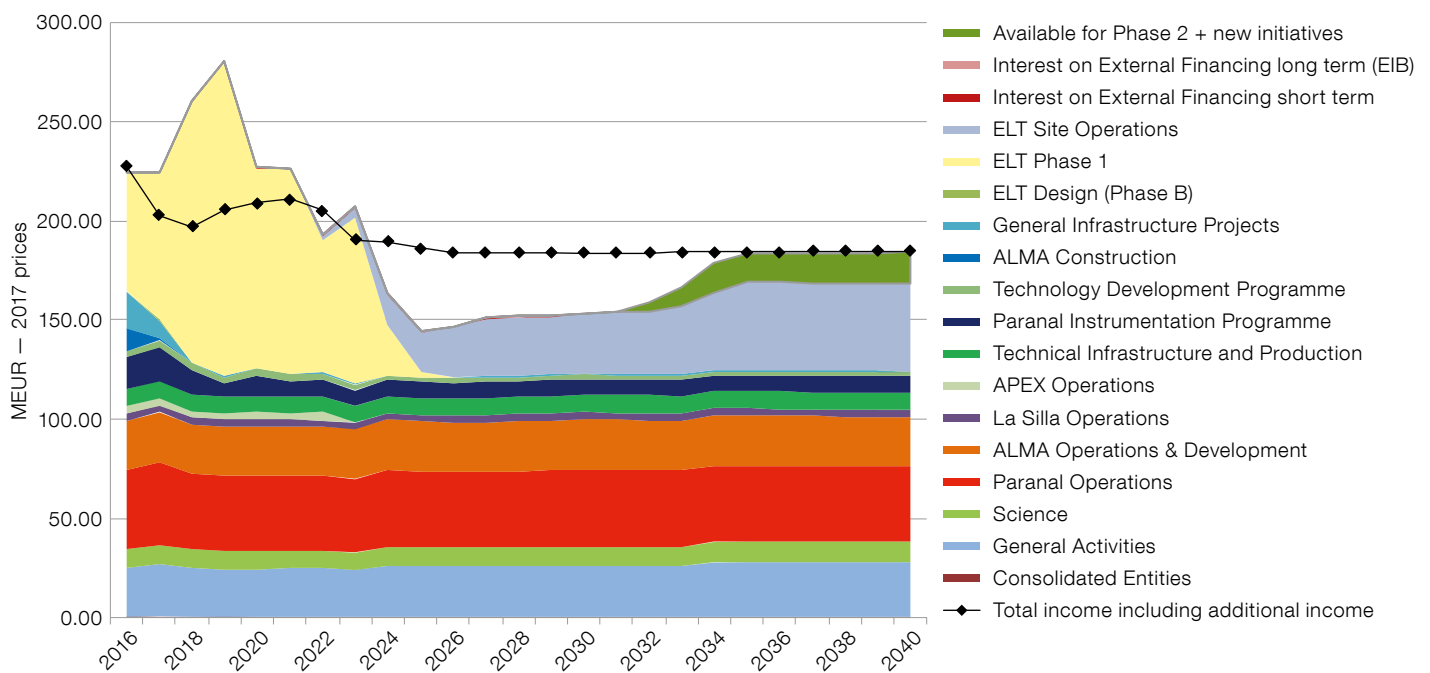

– The interest rates for borrowing and investment are at the European Investment Bank levels.

The assumed values of the exchange rates are based on long-term averages prior to the recent financial crisis. If the euro (EUR) does not recover as expected, then this will increase the pressure on the overall programme. The risk on the US dollar (USD) exchange rate has dropped considerably, as from 2017 onwards the ALMA on-site budget is no longer prepared in USD but in Chilean pesos (CLP). Hedging action in June 2016 made it possible to remove the risk on the CLP exchange rate for the ELT Dome & Telescope Structure contract, as well as for 50 % of the normal operations costs in CLP through 2023.

6.1 Income and expenditure

Figure 16 provides a graphical overview of the income and expenditure in the main budget lines to 2040 (in MEUR, 2017 prices). The regular income consists of the normal annual contributions, the remaining instalments of the special contributions from the Austrian and Polish accessions, and extra income from third parties.

The funding model for ELT construction was developed in 2008, and the specifics were approved by Council in 2011. Since that time, substantial additional costs have been absorbed through a series of measures, including staff reductions and termination of activities, the reduction of the period to calculate the average Swiss Franc (CHF) to EUR exchange rate for the contributions to the Pension Fund, a new Euro-based pension scheme for international staff joining ESO after 1 January 2014, the connection of both La Silla and Paranal to the Chilean electricity grid, and securing the exchange rate of the CLP. Nevertheless, in the baseline plan, building Phase 1 of the ELT with first light in 2024 will require taking up loans to cover the peak expenditure. They will be repaid without having to increase the regular contributions. Advance contributions by the Member States would lower the borrowing burden, and some Member States have already made such contributions.

Phase 2 of the ELT is not funded in the baseline plan. The required 110 MEUR will have to come from new Member States or collaborations with non-Member States, or, if all this were to fail, from the current Member States. Funding of the LTAO unit, which is the highest priority item, could be done by a slight increase in the above mentioned borrowing.

The accession of a new Member State in the near future is not unlikely. Should this occur, it would provide extra resources and hence reduce the risk on the overall programme. It would also lower the borrowing needs during the peak of ELT construction. Depending on the economic size of the new Member State(s), it might be possible to procure some elements of ELT Phase 2.

6.2 Budget lines

The long-term development of the main budget lines is summarised in Figure 16 and the items are summarised below.

*Science* activities continue at the current level with no major changes planned in the coming decade. There is no room for an expansion of the student and fellowship programmes. Education and outreach activities have been increased in order to provide services for the ESO Supernova Planetarium & Visitor Centre. However, the increase will be partly offset by additional income to be generated through fundraising, sponsorship and sales.

The cost of *La Silla Operations* has stabilised owing to the connection to the Chilean power grid and the construction of the photovoltaic plant.

The cost of *Paranal Operations* includes all operations effort in Chile and Garching for the VLT, VLTI, VISTA and VST. Site operations includes expenditure for avoiding obsolescence of critical components. The end-to-end operations remain stable.

Concerning the cost of *APEX Operations* until the end of 2022, ESO's share will increase from 27 % to 32 % in 2018. Additional investments are being made to enable APEX to continue through 2022 (section 3.2).

ESO's share of *ALMA Operations* is expected to remain at a stable level. The expenditure is no longer influenced directly by the EUR/USD exchange rate as the JAO budget is now set in Chilean Pesos. This budget line includes on-site operations by the JAO, off-site operations, the development programme and the ESO ALMA Support Centre.

The *Paranal Instrumentation Programme* is approaching a steady state in the coming years. The approved VLTI programme is nearing completion. The peak of effort associated with the completion of the entire second-generation instrumentation suite (including the AOF and ESPRESSO) will soon be over, allowing a transfer of effort to support the ELT whilst still providing cash and effort to support a healthy ongoing programme. The budget for the latter includes MOONS, ERIS and 4MOST, and will stay flat as of 2021.

The *Technology Development Programme* is set to be flat and will allow continuation of some in-house activities, hopefully matched by funding from external sources and by R&D activities in Member State institutions.

The *Technical Infrastructure and Production* line will also stay flat. This will allow the effective maintenance of the laboratories, workshops and integration halls required to carry out projects. Furthermore the budget will allow small R&D activities, prototyping and hands-on training required to align the skills of the staff to the needs of the ESO programme. This line includes the production of systems (for example detector controllers) funded from external sources.

*General activities* will remain flat over the years despite the expansion of the overall programme with ELT construction. They are subdivided into:
– Garching Headquarters and Santiago operations (maintenance of the various sites, the buildings, electricity, water and heating);
– IT Support and Communication;
– Management and Administration (services provided by the departments of Finance, Contracts & Procurements, and by Human Resources as well as the support of the Director General);
– Governing Bodies and Committees;

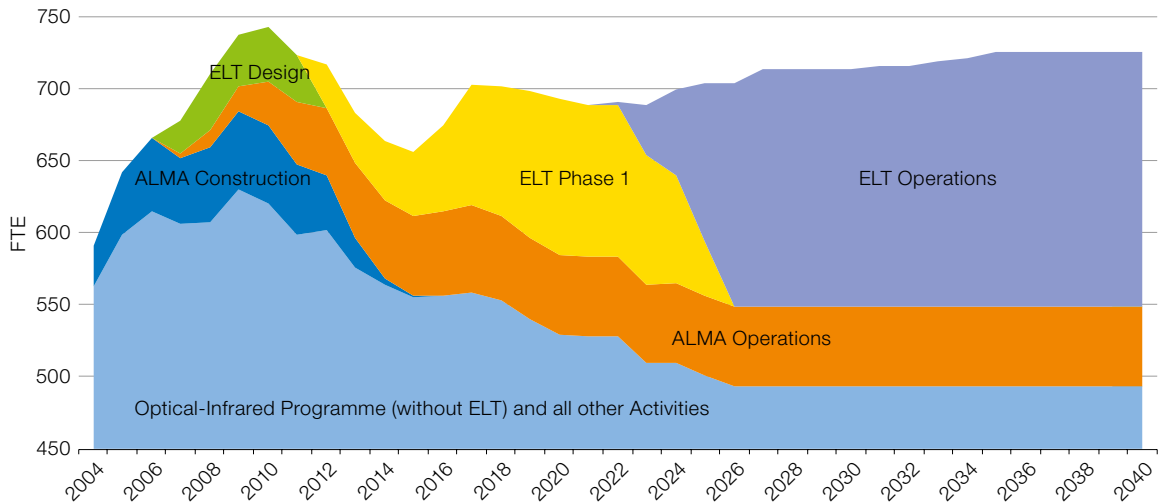

Figure 17. Development of staff complement over the period 2004–2040 focusing on ALMA and ELT. Note that staff numbers begin at 450 FTE.

– Centralised Personnel Budget, containing personnel costs not allocated to projects and activities.

ELT Construction commenced in 2013 with the start of work on the new road to Armazones and the preparation of the platform, together with the procurement of the adaptive M4 mirror. ELT Phase 1 construction is proceeding consistent with first light in 2024 (section 2.2). The spending profile is matched to the procurement schedule. Formally, construction is assumed to continue until 2027 to allow for final payments and the completion of warranty periods.

ELT Operations on-site, including off-site support, will begin in 2022 and ramp up to full operations as of 2027. The upgrade paths, which include the fourth instrument and beyond, will start ramping up from 2030. The slow ramp-up of operations funding and the delayed start of the upgrade paths are required to maintain the financing of Phase 1 with first light in 2024 and to limit the total amount of borrowing.

6.3 Staffing

The staff requirement consists of 440 international staff members and 160 local staff members. The former includes approximately 20 ESO staff members supporting ALMA on-site operations and 19 for APEX (until 2022). Approximately 90 further positions for ESO Fellows, PhD students, and paid and unpaid associates in Garching and Chile support the programmes and their related science. Figure 17 shows the evolution of staff numbers over the period 2004–2040.

7. Looking ahead

ESO's baseline programme supported by the 15 Member States is financially constrained, and does not allow delivery of the Phase 2 components of the ELT in a timely manner. It is therefore useful to summarise developments regarding potential new Member States. This section also considers the relationship with Chile, and describes opportunities in the longer term that are of interest to the astronomical community, and that would be compatible with ESO's mission.

7.1 Brazil and new Member States

The anticipated accession of Brazil has had a major positive influence on ESO, namely the approval of the ELT programme, but the slow progress towards ratification has resulted in challenges for the future, such as the funding of Phase 2 of the ELT, and the availability of sufficient operational funding to allow timely construction of MOSAIC, HIRES and EPICS. Brazil's membership would enable the ELT Phase 2 *in toto*, and would lower the amount of borrowing during ELT construction.

An important milestone for the ratification of the Brazilian Accession Agreement to ESO was reached on 14 May 2015, when the Brazilian Senate approved the ratification proposal. The only remaining steps to conclude the ratification procedure are the signature by the Brazilian President and the deposit of the instrument of accession at the French Ministry of Foreign Affairs. Owing to the political developments in Brazil in 2016, progress on the ratification has stalled. At the time of writing, the final signature by the Brazilian President is still pending. Efforts by Brazilian astronomers and press, supported by ESO, continue in order to further the ratification process.

Informal expressions of interest in a closer association with ESO have been received in past years from many countries, including Armenia, Australia, Canada, China, Hungary, Ireland, Israel, Mexico, Norway, Russia and Turkey. Some of them appear to be moving steadily towards a formal application to join, in particular Hungary, Ireland and Norway. A possible strategic cooperation with Australia is under discussion. Accession of new Member States would provide additional funding and could cover the cost of (part of) the ELT Phase 2, as long as the Council continued the long-standing tradition of adding the annual contribution and the special contribution of new Member States to the existing income.

7.2 Relation with and the role of Chile

ESO decided in 1963 to place its planned observatory in Chile, and the Chilean government was immediately supportive by donating land for the Vitacura premises and for the La Silla Observatory. This was followed by donations for Paranal (1995), for ALMA (2004) and for Armazones (2011). The continued support of the Chilean government, together with the superb natural conditions and the commitment to preserve them, are the main reasons why all ESO observing facilities up to the present are in Chile. The Chilean government has also provided very helpful infrastructure, including paving the public B-710 road that runs through the Paranal area (Figure 7) and enabling the electrical grid connection to Paranal, as well as placing a photovoltaic plant on the La Silla land. It is committed to protect the extraordinary treasure of its clear and dark skies and fosters public appreciation of the night sky as a part of Chile's cultural heritage.

In return, ESO supports a number of activities in Chile, including reserving 10 % of the available observing time for Chilean proposals*. ESO provides funding for postdoctoral fellowships, technology development programmes, outreach and educational activities, conferences and other initiatives which help the growth of Chilean astronomy. In collaboration with Chilean government agencies and the other international observatories, ESO contributes to a joint Office for the Protection of the Quality of the Skies of Northern Chile. About 40 % of ESO's annual budget is spent in Chile, and ESO employs approximately 160 Chilean nationals.

The Chilean astronomical community has grown tremendously in size and quality in the past two decades. There are now astronomy departments at many universities across the country. Reserved access to observing time at world-class facilities has attracted many foreign astronomers to Chilean universities. The quality of proposals from Chilean institutions is nowadays on a par with those from ESO Member States, to the extent that they regularly win more than the reserved share. In the past few years there has also been a push to become involved in the more technical aspects of ESO's activities, including instrument development.

ESO also has an important role in Chile as a leading example of successful intercontinental co-operation between Europe and Latin America. ESO's activities in Chile are a focus of interest for the diplomatic missions and visiting European government representatives, building further support in the current and potential Member States.

The Council invited the Republic of Chile to become a Member State in 1995, but this invitation was declined. The topic was raised again in 2012 by the Chilean President Piñera. Both ESO and Chile have developed considerably over the past two decades, so it is timely to re-examine the arguments.

7.3 New opportunities

The ESO in the 2020s survey (Primas et al., 2015) demonstrated that the astronomical community in the Member States is very happy with what ESO offers, and would like ESO to provide additional facilities. One of these is a large (12–15-metre) optical/infrared telescope dedicated to wide-field spectroscopy which would provide a major step forward in the study of the Milky Way, following up on the Gaia mission, and the deep Universe, complementing imaging data obtained with LSST and the Euclid mission. Innovative designs are being developed. An ESO-led community working group analysed various options in 2015–16 and the conclusions are provided in their report[8]. Such a facility would cost approximately 400 MEUR and might be placed near Paranal, perhaps on La Montura, the mountain behind VISTA. There is interest from various parts of the world, including Australia, Canada and China, in this project, so a partnership could be investigated.

There is also considerable community interest in a large single-dish telescope to complement ALMA, which cannot efficiently study sufficiently large samples of galaxies across wide fields, and will not be able to map degree-scale filamentary Galactic structures in key diagnostic lines. A study, carried out in 2014–15 by another ESO-led community working group, compared various alternatives[9]. It concluded that a 40-metre single dish on the Chajnantor plateau, with a higher-accuracy surface in the inner 25 metres of the dish to enable work at 450 µm, would be the ideal successor to APEX. It would be equipped with panoramic array detectors both for interferometry and for rapid wide-area raster imaging, possibly with a multiplexed direct-detection spectrometer. Such a facility might cost approximately 300 MEUR. Teaming up with one or both of ESO's ALMA Partners should be investigated.

The construction of the ELT will allow the Paranal Observatory to continue to be the centrepiece of ESO's optical/infrared programme well into the 2030s. The STC Paranal White Paper[3] describes possible opportunities for instruments on the UTs driven by technologies that are currently under development. Another area examined was potential new ways of operating the UTs, in the timeframe 2025+, taking into account the plans for the ELT and its instrumentation. For example, pushing high-order adaptive optics into the blue wavelength regime on the UTs, or replicating instruments for the UTs to allow massive surveys on rapid timescales.

Any further expansion of ESO's programme can only come after a reduction of risks on the baseline programme and securing the funding for Phase 2 of the ELT. This expansion should be in line with ESO's mission, and could include participation in one or more of the projects mentioned above. It might also include a further strengthening of the VLTI, with, for example, additional (fixed) ATs and third-generation instrumentation, or an ESO contribution to the cost of site operations of CTA South, or a major upgrade of ALMA beyond the capacity of the development line in the ALMA operations budget.

8. The ESO model

ESO started with five Member States with the goal of building a 3-metre-class telescope in the southern hemisphere. This became the La Silla Observatory in Chile and put the fledgling Organisa-tion on a par with a handful of similar observatories in the world. This was followed by the development of the VLT, ALMA and the ELT, involvement in APEX and now also CTA, accompanied by an expansion to 15 Member States, with more expected to join. The increased programmatic scope has made the Organisation attractive to countries and user communities that would not have considered ESO membership earlier. At the same time, ESO's programmatic expansion would not have been possible without the additional Member States. The two trends — more Member States and programme expansion — are thus inter-related and mutually reinforcing, and have enabled ESO and European astronomy to take the lead in many areas of astronomy. The VLT is the most advanced optical/infrared ground-based facility in the world, ALMA is a transformational radio interferometer and the ELT is on track to be the first and most powerful of the next generation of giant telescopes.

---

\* For the VLT, at least half of the reserved 10 % observing time should be for meritorious proposals in collaboration with astronomers in the ESO Member States; for the ELT this fraction is three-quarters.

8.1 The keys to success

The ESO programme is carried out in strong partnership with the community. This includes joint development of instrumentation, where ESO works closely with institutions in the Member States that provide most of the staff effort in return for guaranteed-time observations. This model was developed soon after 1987 for the VLT instruments and has resulted in a strong and growing network of technical and scientific institutes capable of building very powerful instruments, not only for the VLT but also for the ELT. Further components of the partnership with the community are: an active and attractive student and fellowship programme; a steady flow of small and innovative telescopes and experiments on La Silla, supported by ESO's infrastructure; strong user support of ALMA with a key role for nodes in many of the Member States which are locally funded and provide much additional effort; and the Large Programmes and Public Surveys carried out by international teams who also produce advanced data products adding to the value of the SAF.

Over the past decades, ESO has developed the capability to carry out multiple projects in parallel. The presence of all these activities in one organisation allows: engineering skills to be re-used, benefiting from the lessons learned to work effectively with industry and community partners; testing of new technologies on ESO's own telescopes before applying them to the next-generation projects; and the maximisation of the scientific synergy between the facilities. It has led to significant cross-programme synergy and associated cost savings. This multi-project capability is the result of long-term support by the Member States and can be of value for other ground-based astronomy projects in a cost-effective way, provided the necessary funding is available.

The intergovernmental structure of the Organisation ensures support at the highest level in the Member States and in Chile. This contributes to budget stability and enables the long-term planning ability required for the development of world-leading telescopes. Having all Member States participating in all ESO's programmes is a key additional strength.

The annual contributions are in cash, and set by simple principles. The Organisation is trusted to run an effective industrial competition in the Member States for the procurement of the various components of the telescope systems, to agreed specifications enforced by ESO, aiming at a fair distribution over the Member States. This avoids the significant additional cost and challenges of Juste-Retour** or of one-off projects with a new governance structure and the strong wish that every partner contributes mostly in kind, including for operations.

A further key factor for ESO's development over the past decades is the strong motivation and commitment by all staff to deliver high-quality support and services to ensure the Observatories continue to be world-leading.

---

** Defined in the ESA Convention as: to ensure that all Member States participate in an equitable manner, having regard to their financial contribution.

8.2 Challenges

The health of the overall programme requires full implementation of the ELT funding principles as approved by Council on 6 December 2011. This means that the Member States need to provide the agreed additional contributions on top of regular indexation. Any future non-indexation of the income can only be absorbed by taking a larger loan or by cessation of core activities. Furthermore, it is important to secure the funding for Phase 2 of ELT construction in a timely manner, to reduce the cost of both construction and operations, to increase the scientific return and to bring additional instruments forward.

There are two additional financial risks. The first is that the steady-state operational model for ALMA becomes more expensive than anticipated. The first mitigation step would be to re-discuss with the ALMA partners the level of the off-site development line, which is part of the overall budget line for ALMA. The second risk is the possibility that CERN asks ESO to increase the special organisational contribution to the Pension Fund. It is assumed that the Member States would help ESO in resolving this situation, should it arise.

It is recognised that national budgets in the Member States are under pressure from many directions. Astronomy, and in particular the search for habitable exoplanets, excites many in the general public, and is a good vehicle to attract young people to engineering and the physical sciences, which in turn is critical for the continued well-being of our society. ESO's role in this may be considered an additional argument for continued support.

The multi-project capability developed by the Organisation in the past years must be used with care as it might be interpreted as the capability to do everything at the cost of the quality of the deliverables as well as the motivation of the staff. In addition, it is increasingly difficult to attract sufficiently qualified engineers and technicians from the Member States, as ESO is competing directly with very successful industries. This requires attention in terms of staff development and training and the evolution of the working conditions so as to ensure that ESO continues to be an attractive employer.

Finally, the growth in membership, which has enabled the expansion of the programme including external partnerships, has also increased the complexity of ESO's governance. While national representation is mandatory for Council and Finance Committee, it might be reconsidered for other governing and advisory bodies.

8.3 Strategy for the future

ESO's mission is to (continue to) operate and build world-class facilities enabling astronomical discoveries. This can be done from the ground at optical, infrared and radio wavelengths, and by detection of other messengers from the Universe, be they particles or gravitational waves. ESO's optical/infrared observatories were built by and are operated by ESO. The radio facilities APEX and ALMA are partnerships; ESO operates the former while ALMA is operated by a joint entity created by the ALMA Partnership. Potential new facilities can similarly be "all-ESO" or a partnership. The participation in CTA is an example where ESO will operate the site in Chile on behalf of the CTA Partnership.

It is clear that a moderate further growth in membership would secure Phase 2 of the ELT, and would be necessary for any further increase of the scope of the entire programme. Candidate Member States would be countries with high-quality scientific communities that are keen to join, provide added value, and have government support. ESO currently serves about 30 % of the world's astronomers so there is room for a gradual expansion while keeping a healthy scientific and technical competition with other major observatories.

Whether a new facility would be "all-ESO" or would be developed in a partnership depends on a number of factors. For ALMA the partnership approach was driven by the fact that the power of an interferometer scales proportionally to the square of the number of antennas. Building a larger array with a partner is therefore more cost-effective than building two smaller stand-alone arrays. A partnership may also be preferred when the partners each bring different critical expertise, or when it is the only way to raise the required funding. This could apply to the large single submillimetre dish or to the multi-object spectroscopic telescope mentioned earlier. In other cases, such as a further development of the VLTI, it would be natural to follow the "all-ESO" route.

9. Conclusions

In less than a decade from now, ESO will start operating the ELT, the world's largest optical/infrared telescope and most likely the only one with the ability to image Earth-like exoplanets. The ELT will be an integral part of the Paranal Observatory system, which includes the VLT and VLTI together with a very powerful arsenal of instruments. ALMA will continue to enable tremendous advances in our understanding of the birth of galaxies, stars and planets. Moreover, CTA will have opened a ground-based window on the high-energy Universe to regular observations with ESO operating the southern array in the Paranal area.

This programme implements the Council Resolution on Scientific Strategy of 2004, and in fact goes beyond it by maintaining the VLT as a world-leading instrument well into the next decade and adding CTA South to the programme. This is perfectly in line with ESO's mission, and can be carried out with the support of the 15 Member States, as long as the funding principles agreed by Council are implemented in full. While it may require borrowing during the peak of ELT construction, it will ensure ESO's undisputed leadership in ground-based astronomy into the next decades.

Accession of additional Member States will allow the timely completion of the ELT Phase 2, and with further instruments, which will enhance the scientific power of the ELT, lower its cost and so strengthen the entire programme further. It may also allow ESO to take on another programme soon after ELT construction is completed, and perhaps even a little earlier.

Acknowledgements

Many people contributed in large and small measure to these perspectives. To name them all would be to include a not-insubstantial fraction of ESO staff. In particular the ESO Directors — Rob Ivison (Science), Andreas Kaufer (Operations), Adrian Russell (Programmes), Michèle Peron (Engineering), Patrick Geeraert (Administration) — and the ELT Programme Manager, Roberto Tamai, provided the core of the document. Inputs by Renate Brunner, Mark Casali, Laura Comendador Frutos, Fernando Comerón, Roberto Gilmozzi, Nikolaj Gube, Douglas Pierce-Price, Michael Sterzik, Ewine van Dishoeck, Andrew Williams and Wolfgang Wild were also vital. The Scientific Technical Committee provided helpful feedback on an earlier version. Mafalda Martins and Jeremy Walsh are thanked for layout and editing.

Links

[1] Paranal Instrumentation Programme Plan (Cou-1681): https://www.eso.org/public/about-eso/committees/cou/cou-141st/external/Cou-1681_PIP_6-Month-Report_Sep-2016.pdf
[2] Report from the VLT AO Community Days (STC-581): http://www.eso.org/public/about-eso/committees/stc/stc-88th/public/STC_581_VLT_AO_Community_Days_Report_88th_STC_Meeting.pdf
[3] Paranal in the Era of the ELT (STC-515): https://www.eso.org/public/about-eso/committees/stc/stc-80th/public/STC-515.pdf
[4] Report of the ESO Working Group on Science Data Management (STC-580): https://www.eso.org/public/about-eso/committees/stc/stc-88th/public/STC_580_Data_management_working_group_report_88th_STC_Meeting.pdf
[5] Lessons Learned from ALMA Construction: https://www.eso.org/public/about-eso/committees/cou/cou-135th/external/ESO_ALMA_Construction_Lessons_Learned_public.pdf
[6] High Level Organisational Structure: https://www.eso.org/public/archives/static/about-eso/organisation/eso-high-level-structure-2016.pdf
[7] ESO Science Newsletter: http://www.eso.org/sci/publications/newsletter.html
[8] Report of the Multi-Object Spectroscopy Working Group (STC-579): https://www.eso.org/public/about-eso/committees/stc/stc-88th/public/STC_579_MOS_WG_Report_88th_STC_Meeting.pdf
[9] ESO Sub-millimetre Single Dish Scientific Working Group Report (STC-567): https://www.eso.org/public/about-eso/committees/stc/stc-87th/public/STC-567_ESO_Submm_Single_Dish_Scientific_Strategy_WG_Report_87th_STC_Mtg.pdf

Public ESO Council and STC documents can be found at: https://www.eso.org/public/about-eso/committees/

The Organisation

# Appendix

Resolution approved by Council on 07–08 December 2004

ESO Council, considering the report of its Working Group for Scientific Strategic Planning (Cou 990), and its recommendations, agrees that:

– Astronomy is in a golden age with new technologies and telescopes enabling an impressive series of fundamental discoveries in physics (e.g., dark matter, dark energy, supermassive black-holes, extrasolar planets)
– Over the last decade, the continued investment of ESO and its community into the improvement of ground-based astronomical facilities has finally allowed Europe to reach international competitiveness and leadership in ground-based astronomical research
– The prime goal of ESO is to secure this status by developing powerful facilities in order to enable important scientific discoveries in the future
– Only the continued investment in cutting edge technologies, telescopes, instruments and IT will enable such scientific leadership and discoveries
– ESO will continue to be open to new members and collaborations, following the principle of furthering scientific excellence

and accordingly adopts the following principles for its scientific strategy:

– ESO's highest priority strategic goal must be the European retention of astronomical leadership and excellence into the era of Extremely Large Telescopes by carefully balancing its investment in its most important programmes and projects
– The completion of ALMA is assured and conditions for an efficient exploitation of its superb scientific capabilities will be established
– The VLT will continue to receive effective operational support, regular upgrading (especially to keep it at the forefront in image quality through novel adaptive optics concepts) and efficient 2nd generation instrumentation in order to maintain its world-leading position for at least ten more years
– The unique capabilities of the VLTI will be exploited
– The construction of an Extremely Large Telescope on a competitive time scale will be addressed by radical strategic planning, especially with respect to the development of enabling technologies and the exploration of all options, including seeking additional funds, for fast implementation.

ESO and its community will continue their successful partnership and seek effective intercontinental collaborations in developing the most important and challenging technologies and facilities of the future.

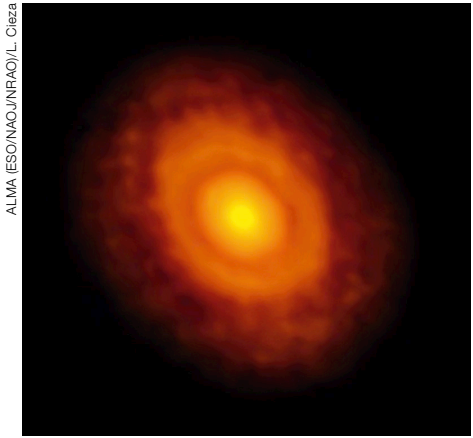
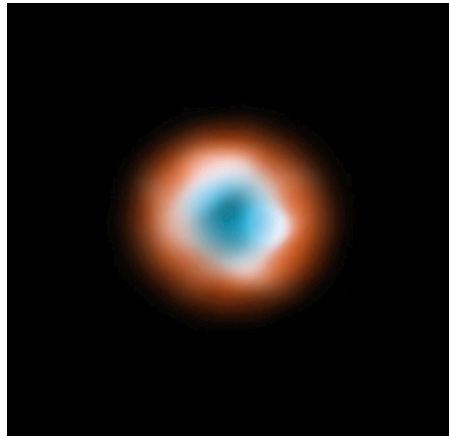
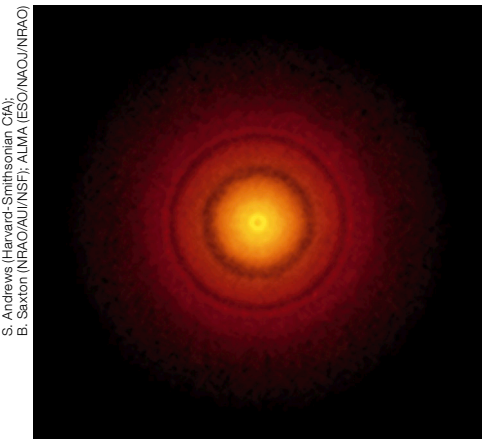
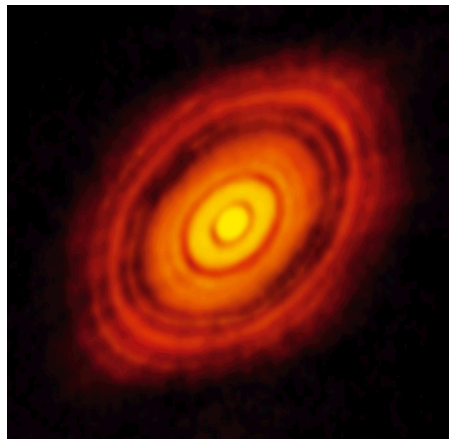

Circumstellar discs with ALMA (left to right, top to bottom): V883 Oronis (1.4 mm); DoAr 44, (870 μm and $^{13}$CO emission); TW Hydrae (870 μm); HL Tauri (1.3 mm).